# Fractional Order AGC for Distributed Energy Resources Using Robust Optimization

Indranil Pan and Saptarshi Das

*Abstract*—**The applicability of fractional order (FO) automatic generation control (AGC) for power system frequency oscillation damping is investigated in this paper, employing distributed energy generation. The hybrid power system employs various autonomous generation systems like wind turbine, solar photovoltaic, diesel engine, fuel-cell and aqua electrolyzer along with other energy storage devices like the battery and flywheel. The controller is placed in a remote location while receiving and sending signals over an unreliable communication network with stochastic delay. The controller parameters are tuned using robust optimization techniques employing different variants of Particle Swarm Optimization (PSO) and are compared with the corresponding optimal solutions. An archival based strategy is used for reducing the number of function evaluations for the robust optimization methods. The solutions obtained through the robust optimization are able to handle higher variation in the controller gains and orders without significant decrease in the system performance. This is desirable from the FO controller implementation point of view, as the design is able to accommodate variations in the system parameter which may result due to the approximation of FO operators, using different realization methods and order of accuracy. Also a comparison is made between the FO and the integer order (IO) controllers to highlight the merits and demerits of each scheme.**

## I. INTRODUCTION

DUE to the deregulation in the energy markets, the environmental emission concerns and the rising costs of electricity transmission and distribution, there is an increasing trend of shifting from centralized power generation and distribution to a more decentralized mode [1]. This has given rise to distributed energy resources (DERs) with integration of renewable energy technologies like wind and solar, along with energy storage devices like flywheels, batteries etc. and combined heat and power generation technologies [1]. Control and communication play an important role in the efficient operation of these distributed power systems [2]. To effectively meet the challenges of control in DERs, the paper looks at a novel controller design strategy for a hybrid power system [3], where the sensor measurements and the control signals are sent over an unreliable communication network introducing random time delays in the control loop [4].

Fractional calculus [5] is a 300 year old mathematical concept. However, in the last decade, it has found applications in control systems and is gaining increasing interest from the research community in other domains as well. Fractional calculus has also found recent applications in computational intelligence based control system design [6] with random time delays e.g. in process control [7], nuclear reactor control [8] etc. among others. Inspired by the successful applications in these domains, the present paper investigates the applicability of fractional order intelligent control for hybrid power system or distributed energy generation. Other control approaches of similar kind of system design include the standard PID controller [9], robust H∞ controller [10] etc.

The stochastic nature of the demand load and the renewable generation terms i.e. solar and wind energies introduces fluctuations in the system frequency [11-12]. The controller in the hybrid power system tries to minimize the aberrations in the system frequency so that the power quality is maintained. This leads to the concept of AGC for grid frequency oscillation damping in the context of distributed energy generation [13]. This is done by sending an appropriate control signal to the energy storage systems to absorb (release) the surplus (deficit) power from (to) the grid. The controllers are generally tuned in an output optimal fashion by minimizing some error criterion [9]. However, these obtained values of controller parameters may not be robust to the slight variability during actual hardware implementation. This is more relevant for fractional order controllers, since they are realized in hardware using band limited approximations of higher order transfer functions using different techniques like Carlson, Matsuda, Continued Fraction Expansions (CFE) etc. [5, 6, 14]. Under such circumstances, the optimal response as obtained in the time domain simulations would vary significantly. Therefore a robust optimization based controller parameter tuning scheme is proposed in the present paper to overcome this issue and facilitate practical implementation. The controllers tuned with the robust algorithms show slight variations in time domain performance, as opposed to the drastic deterioration of performance with the optimally tuned controllers, when the controller parameters are perturbed. In the present work, different variants of PSO are used for robust optimization. PSO has been used in other smart grid applications like demand response and resource scheduling [15], sizing of distributed generation and storage capacity [16], wind power control [17], bi-directional energy trading [18] etc. The present paper introduces a FOPID based centralized AGC scheme for grid frequency oscillation damping in a DER system. We use the PSO based robust optimization technique for tuning the controller and report the achievable parametric robustness of the hybrid power system.

## II. DESCRIPTION OF THE DISTRIBUTED ENERGY SYSTEM

The schematic representation of the hybrid power system using different energy generation/storage is illustrated in Fig. 1 with its various components described in Table I.

### A. Models of Different Generation Subsystems

TABLE I
NOMINAL PARAMETERS OF THE COMPONENTS OF HYBRID POWER SYSTEM

| Component | Gain ($K$) | Time constant ($T$) |
|---|---|---|
| Wind turbine generator (WTG) | $K_{WTG} = 1$ | $T_{WTG} = 1.5$ |
| Aqua Electrolyzer (AE) | $K_{AE} = 0.002$ | $T_{AE} = 0.5$ |
| Fuel Cell (FC) | $K_{FC} = 0.01$ | $T_{FC} = 4$ |
| Flywheel energy storage system (FESS) | $K_{FESS} = -0.01$ | $T_{FESS} = 0.1$ |
| Battery energy storage system (BESS) | $K_{BESS} = -0.003$ | $T_{BESS} = 0.1$ |
| Diesel engine generator (DEG) | $K_{DEG} = 0.003$ | $T_{DEG} = 2$ |
| Solar Photovoltaic (PV) | $K_{PV} = 1$ | $T_{PV} = 1.8$ |

For small signal analysis, the dynamics of the WTG, PV, FC and DEG can be modeled by first order transfer functions (1)-(4) with the associated gain and time constants given in Table I [9, 19], where $k$ represents the number of units. These transfer functions represent the electrical power produced ($P_{WTG}$, $P_{PV}$) from the renewable energy sources like wind power ($P_w$), solar irradiation ($\Phi$) etc. In this study, a centralized controller has been used for the hybrid energy system (Fig. 1) as opposed to multiple decentralized controllers [9] for each individual sub-systems like the battery, flywheel and diesel. This helps in easier maintenance, reduced wiring and also makes the design problem tractable by reducing the number of controller parameters. However, there would obviously be some deterioration in the performance, as the same control signal is being used for all the sub-systems. Nevertheless, here we show that the centralized controller results in acceptable time domain performance. Different rate limiters in each subsystem have been provided so that the control signal is appropriately modified with respect to the individual electromechanical characteristics of the storage/generating devices.

$$G_{WTG_k}(s) = K_{WTG}/(1+sT_{WTG}) = \Delta P_{WTG}/\Delta P_W, k=1,2,3 \quad (1)$$

$$G_{PV}(s) = K_{PV}/(T_{PV}s+1) = \Delta P_{PV}/\Delta \Phi \quad (2)$$

$$G_{FC_k}(s) = K_{FC}/(1+sT_{FC}) = \Delta P_{FC_k}/\Delta P_{AE}, k=1,2 \quad (3)$$

$$G_{DEG}(s) = K_{DEG}/(1+sT_{DEG}) = \Delta P_{DEG}/\Delta \tilde{u} \quad (4)$$

### B. Model of Aqua Electrolyzer

The aqua-electrolyzer produces hydrogen for the fuel cell by using a part of the power generated from the renewable sources like wind and/or solar. The dynamics of the AE can be represented by the transfer function (5) [19] and it uses $(1-K_n)$ fraction of the total power of WTG and PV to produce hydrogen which is again used by two FCs to produce power as an additional source to the grid.

$$G_{AE}(s) = K_{AE}/(1+sT_{AE}) = \Delta P_{AE}/(\Delta P_{WTG} + \Delta P_{PV})(1-K_n) \quad (5)$$

where, $K_n = P_t/(P_{WTG}+P_{PV}), K_n = 0.6$.

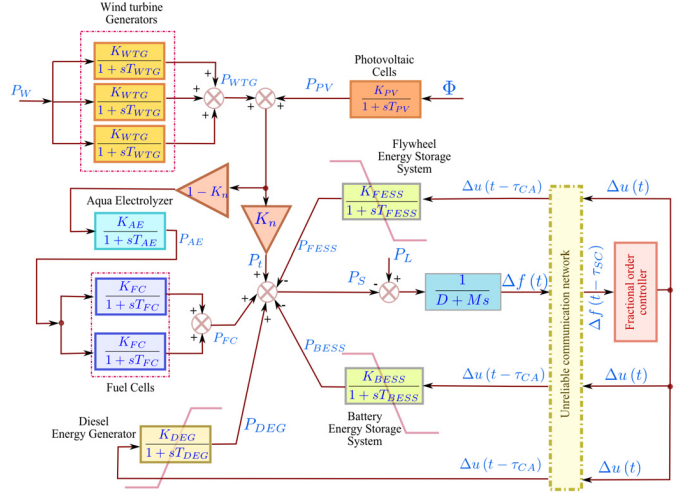

Fig. 1 Schematic of the hybrid power system used in this study.

### C. Models of Different Energy Storage Systems

In the hybrid energy system of Fig. 1, the FESS and the BESS are connected in the feedback loop and are actuated by the signal from the FOPID or PI$^\lambda$D$^\mu$ controller. These absorb or release energy from or to the grid if there is a surplus or deficit amount of power respectively. Their corresponding dynamical models can be represented by (6)-(7) [19].

$$G_{FESS}(s) = K_{FESS}/(1+sT_{FESS}) = \Delta P_{FESS}/\Delta \tilde{u} \quad (6)$$

$$G_{BESS}(s) = K_{BESS}/(1+sT_{BESS}) = \Delta P_{BESS}/\Delta \tilde{u} \quad (7)$$

Here, the incremental control action of the FOPID controller is represented as $\Delta \tilde{u}(t) = \Delta u(t-\tau_{CA})$, $\tau_{CA} \sim U(0.05, 0.15)$ [7] that actuates the energy storage/generation devices. The controller output $\Delta u$ gets corrupted by the stochastic network delay ($\tau_{CA}$) for the transmission of the signal from the controller to actuator. Also, the models of the grid frequency dependent energy storage/generating elements are considered to have rate constraint nonlinearities as $|\dot{P}_{FESS}| < 0.9$, $|\dot{P}_{BESS}| < 0.2$, $|\dot{P}_{DEG}| < 0.01$ respectively (Fig. 1), such that all the three components operate in the nonlinear zone for a wide range of controller values. The rate constraint nonlinearities take care of the various electromechanical constraints that these devices exhibit. However, the overall system dynamics is dictated by the relative values of the gains and time constants of the different components. For example, the DEG has a larger time constant as compared to the FESS. Therefore, as soon as a control signal is applied, the FESS will respond more quickly and the DEG would take more time to respond. Therefore, the overall system dynamics would be governed by a combination of these fast and slow dynamics.

### D. Power System Model Using Grid Frequency Deviation

The power system model can be represented as (8) which describes the dynamics of power deficit/surplus ($\Delta P_e$) to the grid frequency oscillation ($\Delta f$).

$$G_{sys}(s) = \Delta f / \Delta P_e = \Delta f / (P_L - P_S) = 1/(D + Ms) \quad (8)$$

where, $M$ and $D$ are the equivalent inertia constant and damping constant of the hybrid power/energy system [3] and their typical values are considered as 0.4 and 0.03 respectively for the present simulation study.

### E. Wind Speed Model

The wind speed model should be able to capture the spatial dependencies of wind flow like base fluctuation and small stochastic components etc. To achieve this objective, the two component wind model [19] is chosen as represented by (9).

$$V_W = V_{WB} + V_{WN} \quad (9)$$

where, $\{V_{WB}, V_{WN}\}$ are the base wind component and noise wind component respectively. The base wind component is the constant component which is always present during the operation of the wind turbine and has been considered as (10).

$$V_{WB} = K_B \left[ 7.5H(t) - 3H(t - 200) + 10.5H(t - 250) \right] \quad (10)$$

where, $K_B$ is a constant (within a constant speed operation regime) and $H(t)$ represents the Heaviside step function. The noise component of the wind is expressed as (11).

$$V_{WN} = 2\sigma^2 \sum_{i=1}^{N} \sqrt{S_V(\omega_i)\Delta\omega} \cos(\omega_i t + \varphi_i) \quad (11)$$

where, $\omega_i = (i - 1/2)\Delta\omega$, $\varphi_i \sim U(0, 2\pi)$, $\Delta\omega$ is the frequency step to compute spectral density, $\sigma^2 = 200$ is variance of the noise component and the spectral density function $S_V(\omega_i)$ is given by (12).

$$S_V(\omega_i) = 2K_N F^2 |\omega_i| \Big/ \left( \pi^2 \left[ 1 + (F\omega_i / \mu\pi)^2 \right]^{4/3} \right) \quad (12)$$

Here, $K_N = 0.004$ is the surface drag coefficient, $F = 2000$ is the turbulence scale and $\mu = 7.5$ is the mean wind speed at reference height. Here, $N = 50$ and $\Delta\omega = 0.5$ rad/s are taken to achieve an effective modeling accuracy.

### F. Characteristics of Wind Turbine Model Output

The non-dimensional curves of the power coefficient $C_p$ expressed as a function of tip speed ratio $\lambda$ (14) and blade pitch angle $\beta = 0.1745$ is used to characterize the wind turbine and is expressed as (13) [19].

$$C_p = (0.44 - 0.0167\beta) \sin\left[ \frac{\pi(\lambda - 3)}{15 - 0.3\beta} \right] - 0.0184(\lambda - 3)\beta \quad (13)$$

Here, $\lambda$ refers to the ratio of the speed at the blade tip of the wind turbine to the wind speed and is expressed as (14).

$$\lambda = R_{blade}\omega_{blade} / V_W \quad (14)$$

where, $R_{blade} = 23.5$m is the radius of the wind turbine blades and $\omega_{blade} = 3.14$ rad/s is the rotational speed of the blades.

The output mechanical power of the wind turbine is given by

$$P_W = (1/2)\rho A_r C_p V_W^3 \quad (15)$$

where, $\rho = 1.25$ kg/m$^3$ refers to the air density and $A_r = 1735$ m$^2$ is the swept area of the blades.

### G. Characteristics of PV Output Power and Demand Power

The power output of the photovoltaic system can be represented by (16) as also done in [19].

$$P_{PV} = \eta S \Phi \left\{ 1 - 0.005(T_a + 25) \right\} \quad (16)$$

where, $\eta = 10\%$ is the conversion efficiency of the PV cells, $S = 4084$ m$^2$ is the measured area of the PV array, $\Phi$ (17) in kW/m$^2$ is the solar radiation on the surface of the PV cells and $T_a = 25$ °C is the ambient temperature.

$$\Phi = 0.5H(t) - 0.3H(t - 25) + 0.3H(t - 75) - 0.3H(t - 150) + \Phi_n(t), \Phi_n(t) \sim U(-0.1, 0.1) \quad (17)$$

Fig. 2 depicts one realization of the stochastic generation components ($P_{PV}, P_{WTG}$) along with the stochastic variations in the load demand $P_L$ (18) and the net power generated by the renewable sources to the grid ($P_t$).

$$P_L = H(t) - 0.4H(t - 50) - 0.1H(t - 100) + 0.2H(t - 175) + 0.2H(t - 225) + N_L, \quad N_L \sim U(-0.05, 0.05) \quad (18)$$

In all the three cases, there are sudden fluctuations in the power levels with stochastic aberrations throughout in $P_{WTG}$, $P_{PV}$ and $P_L$ which is representative of a realistic scenario.

**Stochastic fluctuation of the power generated by renewable sources and demand load**

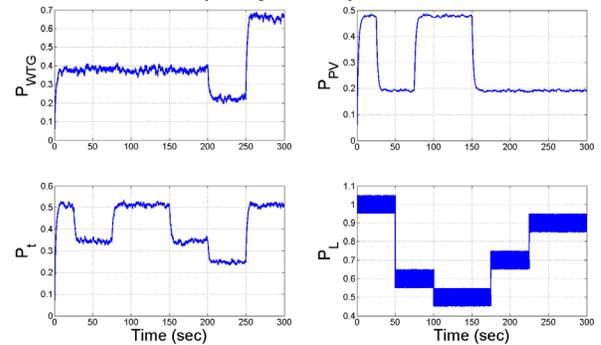

Fig. 2 A single realization of the renewable generations and demand powers which are independent of the controller structure.

### H. Control Over Unreliable Communication Network

Since the different energy resources are located at different places, they are assumed to communicate *via* a shared communication medium [20, 21]. The use of a shared medium introduces random delays in the control loop between the grid frequency sensor to the controller ($\tau_{SC}$) and the controller to the actuator ($\tau_{CA}$) [22]. A small amount of random delay can induce instability while the control loop can be stable with a larger amount of lumped static delay [6, 7]. Therefore these stochastic delays must be considered in the optimization based controller design procedure itself [8]. In the present simulation, $\tau_{SC}$ and $\tau_{CA}$ are assumed to be randomly drawn from a uniform distribution within a range of $[0.05, 0.15]$. The incremental control signal before and after the network is represented as $\Delta u$ and $\Delta \tilde{u}$. The whole hybrid power system model in Fig. 2 thus can be viewed as a higher order stochastic delay differential equation due to stochastic forcing and stochastic delay terms and is numerically integrated with the 3rd order accurate Bogacki-Shampine formula over a total time window of 300 sec with a fixed step-size of 0.01 sec.

## III. BASICS OF FRACTIONAL ORDER CONTROLLER

Fractional calculus is an extension of the $n^{th}$ order successive differ-integration of a function $f(t)$ having the order as any real value $n = \alpha$. For control system studies, the non-integer order integro-differential operator is defined as Caputo derivative (19) among three main definitions of fractional calculus which under zero initial condition in Laplace transform produce FO transfer functions [5].

$$_0 D_t^\alpha f(t) = \frac{1}{\Gamma(m-\alpha)} \int_0^t \frac{D^m f(t)}{(t-\tau)^{\alpha+1-m}} d\tau,$$
$$\alpha \in \Re^+, m \in Z^+, m-1 \le \alpha < m$$
(19)

The transfer function representation of a FOPID controller is given in (20).

$$C(s) = K_p + \left(K_i / s^\lambda\right) + K_d s^\mu$$
(20)

This typical controller structure has five independent tuning knobs i.e. the three controller gains $\{K_p, K_i, K_d\}$ and two FO integro-differential operators $\{\lambda, \mu\}$. For $\lambda = 1$ and $\mu = 1$ the controller structure (20) reduces to the classical PID controller in parallel structure [14, 6].

Various continuous and discrete time rational approximation methods can be adopted to implement the FO operators. In this paper, each guess value of the FO differ-integrals $\{\lambda, \mu\}$ within the optimization process is continuously rationalized with the Oustaloup's 5th order rational approximation (ORA) [14] within the chosen frequency range of $\omega \in \{10^{-2}, 10^2\}$ rad/sec. Due to the fact that FO differ-integrals represent infinite dimensional linear filters, their band-limited realizations are necessary for implementation. Here, each FO element has been rationalized using ORA given by the equations (21) and (22). If it be assumed that the expected fitting range or frequency range of controller operation is $(\omega_b, \omega_h)$, then the higher order filter which approximates the FO element $s^\gamma$ can be written as (21).

$$G_f(s) = s^\gamma \approx K \prod_{k=-N}^{N} \frac{s + \omega'_k}{s + \omega_k}$$
(21)

The poles, zeros and gain of the filter can be evaluated as (22).

$$\omega_k = \omega_b \left(\omega_h / \omega_b\right)^{\frac{k+n+0.5(1+\gamma)}{2n+1}}, \omega'_k = \omega_b \left(\omega_h / \omega_b\right)^{\frac{k+n+0.5(1-\gamma)}{2n+1}}, K = \omega_h^\gamma$$
(22)

In equations (21)-(22), $\gamma$ is the order of the differ-integration and $(2n+1)$ is the order of the realized analog filter. The controller operates on the randomly delayed grid frequency deviation signal to produce the control action (23).

$$u(t) = \left(K_p + K_i D^{-\lambda} + K_d D^\mu\right) \Delta f(t - \tau_{SC}), \tau_{SC} \sim [0.05, 0.15]$$
(23)

## IV. OPTIMIZATION ALGORITHMS AND CONTROL OBJECTIVES

### A. The Concept of Robust Optimization

The difference between a robust and an optimal solution is illustrated in Fig. 3. The optimum point has the lowest value of the objective function. However, if the input design variable has a certain variance, indicated by the first probability distribution curve on the abscissa, then there is a corresponding large deviation in the objective function value, indicated by the probability distribution curve on the ordinate. In case of the robust solution, the same variance in the input variable produces a smaller variance in the objective function value. Hence the latter solution is less sensitive to variation in system parameters and is consequently a robust solution. The robust solution has a higher value of objective function than the optimal solution, but the worst case scenario for the robust solution is much less severe than that of the corresponding optimum solution [23].

There are various methods of assessing the robustness of solutions $(\vec{x})$ for the objective function $J(\vec{x})$. The expected fitness measure is used in this paper and is given by (24).

$$J_{\exp}(\vec{x}) = \int_{-\infty}^{\infty} J(\vec{x} + \vec{\delta}) . pdf(\vec{\delta}) d\vec{\delta}$$
(24)

where, $\vec{\delta}$ is the input variable fluctuations, $pdf(\vec{\delta})$ is the probability distribution function of the occurrence of $\vec{\delta}$ over the whole input variable space $[-\infty, \infty]^N$, and $N$ is the problem dimension i.e. the number of decision variables.

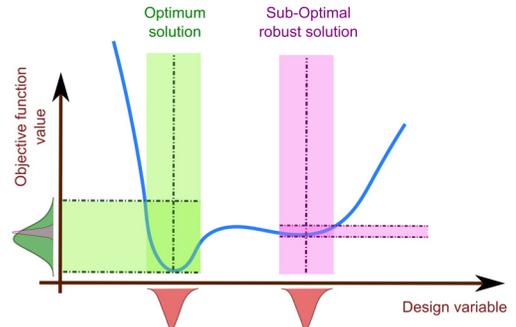

Fig. 3 Schematic showing the difference between an optimum solution and a robust solution.

### B. Objective Function for Optimization Based Control

For effective functioning of the hybrid power system, the controller gains and FO integro-differential orders need to be tuned. For the controller design problem, the objective function in (25) is considered. It consists of the integrals of two weighted terms, which try to minimize the frequency deviation in the hybrid power system ($\Delta f$), as well as the incremental control signal ($\Delta u$) [7-8, 24].

$$J = \int_0^{T_{max}} \left[ w(\Delta f)^2 + \left((1-w)/K_n\right)(\Delta u)^2 \right] dt \quad (25)$$

where, $w$ dictates the relative importance of the two objectives (i.e. Integral of Squared Error – ISE and Integral of Squared Deviation of Control Output – ISDCO) and $w = 0.5$ is considered to give equal weightage on both the parts of the control objective. $K_n = 10^5$ is the normalizing constant to scale ISE and ISDCO in an uniform scale and $T_{max} = 300$ s .

The objective function (25) is formulated in such a way that along with the frequency oscillation ($\Delta f$), the incremental control signals going to different actuators are also minimum. This helps to limit the requirement of increased capacity for the battery, reduces flywheel jerk and diesel consumption, making the overall hybrid power system more cost-effective.

Also, due to the noisy nature of the frequency deviation signal, the control signal gets amplified with the derivative action of the PID/FOPID and even more due to the difference operator ($\Delta \underline{u}$) of the control signal. Therefore, there is a need of a scale factor to bring the amplitude of the ISE and ISDCO to a comparable platform and then only appropriate weights could be assigned by a designer. An equal weightage is given in the present case, as also reported in [7-8, 24]. In most of the controller design problems, it is difficult to decide the weighs *a priori* and a multi-objective optimization formalism should be used to obtain the Pareto optimal trade-offs for different weights as shown in [25]. The designer can then choose the suitable weighting according to his requirements. Nevertheless, the proposed methodology is still valid if the weightings in the objective function is changed and our simulations show one of such possible alternatives.

### C. Implementation Issues of Fractional Order Controller and Need for Robust Optimization

Depending on the different methods of realization i.e. analog (like Crone, Carlson, Matsuda, CFE with high and low frequency approximations etc.) and digital (Tustin, Simpson, backward difference, impulse response etc.) methods of a FO element, the time/frequency domain characteristics of the filter may be different [14]. Fig. 4 shows the phase response of the band limited realization of $1/\sqrt{s}$ in the frequency range of $10^{-2} - 10^2$ Hz. The corresponding time domain impulse responses are shown in Fig. 5. It can be seen that there exists significant differences in the time domain response among the different realizations and also among different approximation orders of the same realization. From the application point of view, one can opt for any one of these realization techniques and the filter order to implement a single FO operator without paying much attention to the resulting frequency and time domain discrepancies of the approximation. Therefore, the design of the controller parameters themselves should be robust enough to tolerate these imprecisions during the actual hardware implementation, while still ensuring satisfactory time domain performance. In addition, the components of the hybrid energy system are generally modelled as low-pass first order transfer functions, considering small signal stability analysis [19], whereas in reality they may show more complex nonlinear behavior which can be considered as a linear model with uncertain parameters. The concept of robust optimization in the present problem is introduced to handle both system nonlinearities and FO controller implementation issues.

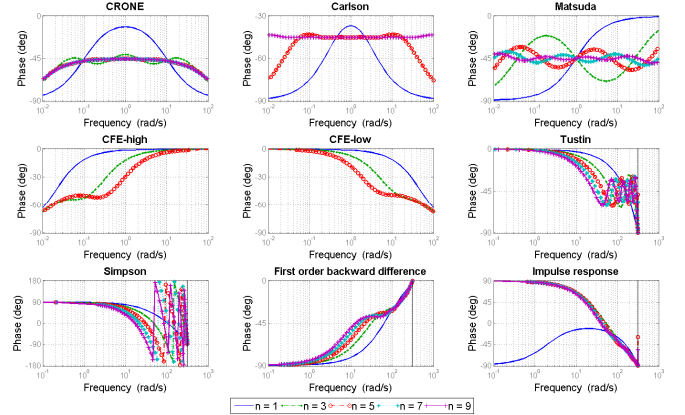

Fig. 4 Phase responses of different band-limited realizations of FO element $1/\sqrt{s}$ with different methods and order of approximation.

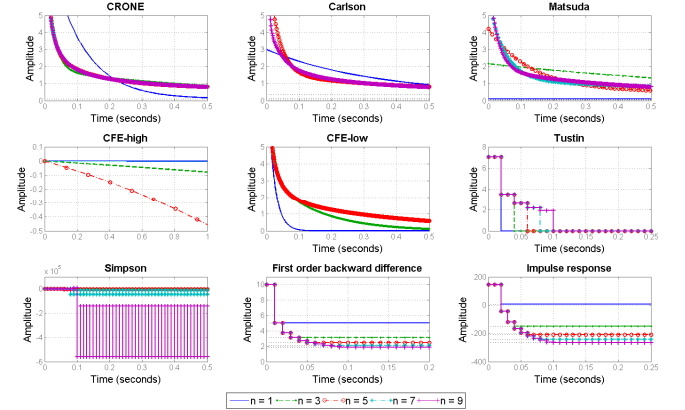

Fig. 5 Impulse responses of FO element $1/\sqrt{s}$ with different methods/orders.

### D. Canonical PSO (CPSO) Optimizer and Its Variants

The CPSO algorithm tries to optimize an objective function $J\left(\vec{x}\right)$ with respect to the design variable $x \in \Re^n$ as in (26).

$$\underset{x \in \Re^n}{\text{minimize}} J\left(\vec{x}\right) \quad (26)$$

where, the objective function $f : \Re^n \to \Re$ and the $n$-dimensional search space $G \in \Re^n$ is pre-specified by the user. The PSO algorithm consists of a swarm of particles $x_i \; \forall \; i \in \{1, 2, ..., n_p\}$ with the maximum number of particles $n_p$ specified by the user. The particles $x_i$ search for an optimal solution $x' \in \Re^n$ of (26). The position of the $i^{th}$ particle is denoted by $x_i := \left(x_{i,1}, x_{i,2}, ..., x_{i,n}\right)^T \in \Re^n$ and the velocity is denoted by $v_i := \left(v_{i,1}, v_{i,2}, ..., v_{i,n}\right)^T \in \Re^n$, where $i \in \{1, 2, ..., n_p\}$. The position and velocity of the $i^{th}$ particle, $x_i \in \Re^n$ is updated in each iteration, based on equations (27)-(28) for $k \in Z_+$ which indicates the iteration number.

$$x_i^{k+1} = x_i^k + v_i^{k+1} \quad (27)$$

$$v_i^{k+1} = \chi\left[v_i^k + \beta_1\theta_{1,i}^k\left(x_i^{best,k} - x_i^k\right) + \beta_2\theta_{2,i}^k\left(x_{nbr}^{best,k} - x_i^k\right)\right] \quad (28)$$

where, $\chi$ is the constriction coefficient, $\beta_1$ is the cognitive learning rate and $\beta_2$ is the social learning rate and are specified by the user and $\{\theta_{1,i}^k, \theta_{2,i}^k\} \sim U(0,1)$ are random variables. $x_i^{best,k}$ in (28) refers to the previously obtained best position of the $i^{th}$ particle and $x_{nbr}^{best,k}$ denotes the best position of the particle neighbors at the current iteration $k$ as in (29).

$$x_i^{best,k} := \arg\min_{x_i^j}\left\{J(x_i^j), 0 \le j \le k\right\} \quad (29)$$

$x_{nbr}^{best,k}$ is dependent on the topology of the swarm and for a fully connected topology it can be expressed as (30).

$$x_{nbr}^{best,k} := \arg\min_{x_i^k}\left\{J(x_i^k), \forall i\right\} \quad (30)$$

Another variant of the PSO, known as the Fully Informed Particle Swarm (FIPS) is almost identical to the CPSO, but differs only in the velocity update equation.

$$v_i^{k+1} = \chi\left[v_i^k + (1/\Lambda_i)\sum_{m=1}^{\Lambda_i}\beta_1\theta_{1,i}^k\left(x_{nbr,m}^k - x_i^k\right)\right] \quad (31)$$

where, $\Lambda_i$ is the number of neighbors of particle $i$ and $x_{nbr,m}^k$ is the $m^{th}$ neighbor of particle $i$. In (31) all the solutions are taken into account instead of just the best neighbor.

A second variant of the CPSO algorithm is the Charged CPSO (CCPSO) algorithm differs in the velocity update equation and is given by (32).

$$v_i^{k+1} = \chi\left[v_i^k + \beta_1\theta_{1,i}^k\left(x_i^{best,k} - x_i^k\right) + \beta_2\theta_2^k\left(x_{nbr}^{best,k} - x_i^k\right) + \alpha_i^k\right] \quad (32)$$

where, $\alpha$ is the acceleration/repelling coefficient. This helps in maintaining the diversity of the population and avoids premature convergence. The value of $\alpha_i^k$ is calculated as

$$\alpha_i^k = \sum_{l=1, l\neq i}^{n_p}\alpha_{il}^k \quad (33)$$

where, $\alpha_{il}^k$ is the repulsion force between particle $i$ and particle $l$ at iteration $k$ and is calculated as (34).

$$\alpha_{il}^k = \begin{cases} Q_{il}\left(x_i^k - x_l^k\right)/\left\|x_i^k - x_l^k\right\|^3, & R_c \le \left\|x_i^k - x_l^k\right\| \le R_p \\ Q_{il}\left(x_i^k - x_l^k\right)/R_c^2\left\|x_i^k - x_l^k\right\|^3, & R_c > \left\|x_i^k - x_l^k\right\| \\ 0, & R_p < \left\|x_i^k - x_l^k\right\| \end{cases} \quad (34)$$

where, $Q_{il}$ is the charge value between particle $i$ and $l$, $R_c$ is the core radius and $R_p$ is the perception limit of each particle.

The termination criterion for the PSO variants are set as the user specified maximum number effective fitness function evaluations (=2500). The population is taken as 30 and the values of $\beta_1$ and $\beta_2$ are chosen as 2.8 and 1.3 respectively for all the variants. A ring topology is adopted for the CPSO algorithm as it ensures higher diversity and helps to find better global minima. For CCPSO, $Q_{il} = 1, R_c = 1, R_p = 2\times\max(\Delta)$, where $\Delta$ is the perturbation of the controller gains and orders.

## E. Archive Strategy in PSO Variants for Robust Optimization

The expected value of the robust objective function in (24), can be approximated by sampling the perturbed objective function multiple times and taking the mean of the result. This can be expressed as (35) and is known as the multi-evaluation model [26].

$$\hat{J}_{exp}\left(\vec{x}\right) = \frac{1}{n}\sum_{i=1}^{n}f\left(\vec{x} + \vec{\delta_i}\right) \quad (35)$$

where, $\vec{\delta_i} \sim pdf\left(\vec{\delta}\right)$ and $n$ is the number of function evaluations. It is evident that increasing $n$ would result in a better approximation but would be more computationally expensive. To decrease the number of function calls to the original objective function, an archive is created [27, 28] which can store the points $\vec{x}$ and the corresponding fitness values $J\left(\vec{x}\right)$. Samples are added to the archive based on a selection algorithm [27, 28]. For reading the samples from the archive a Latin Hypercube sampling scheme is created and the corresponding closest representative points in the archive are calculated. If the representative archive point is also closest to the candidate sampling point, it is added to the set $S$, otherwise, it is added to a resampling candidate point set $X$. The set of points $S$ can be used for robust fitness evaluation without calling the computationally expensive objective function, while those in $X$ need to be resampled and added to the archive. The effective fitness function is calculated by using the points from both $S$ and $X$ using (36).

$$\hat{J}_{exp}\left(\vec{x}\right) = \left(1/|S \cup X|\right)\sum_{\vec{x'} \in X \cup S}f\left(\vec{x'}\right) \quad (36)$$

More accurate approximation can be obtained by utilizing all the available and usable values in the archive $A_u$ and (36) can be modified to (37) [29].

$$J_{eff}\left(\vec{x}\right) = \sum_{\vec{x'} \in X \cup A_u}w\left(\vec{x'}\right)\cdot f\left(\vec{x'}\right) \Bigg/ \left(\sum_{\vec{x'} \in X \cup A_u}w\left(\vec{x'}\right)\right) \quad (37)$$

where, $A_u$ are all the archive points from the area of interest and $w\left(x'\right) \sim pdf\left(\vec{\delta}\right)$. If the uncertainty in the input parameters is a uniform distribution (as in the present case), then the weighting function can be expressed as

$$w\left(\vec{x'}\right) = \begin{cases} 1, & if \ \vec{x'} \in \left[\vec{x} - \vec{\delta}, \vec{x} + \vec{\delta}\right] \\ 0, & otherwise \end{cases} \quad (38)$$

and the corresponding effective fitness function in (37) reduces to (39).

$$J_{eff}\left(\vec{x}\right) = \frac{1}{|X \cup S \cup A_u|}\sum_{\vec{x'} \in X \cup A_u}f\left(\vec{x'}\right) \quad (39)$$

An archive cleanup procedure, as outlined in [28] is also used and the maximum number of elements in the archive is limited to 5000. The number of samples for effective fitness approximation is taken as 10.

## V. RESULTS AND DISCUSSIONS

### A. Controller Design for Distributed Energy Resources

The optimal and robust PSO based optimization algorithms [28, 30] as discussed in Section IV are employed to optimize the distributed power/energy system as described in Section II. For robust tuning, the perturbation of the controller gains and orders from their nominal values are considered as $\Delta\{K_p, K_i, K_d, \lambda, \mu\} \in [10, 10, 0.15, 0.2, 0.2]$ (in both positive and negative directions). The optimal PID/FOPID controllers are tuned using the CPSO algorithm and the robust PID/FOPID controllers are tuned with the three variants of the PSO algorithm, viz. CPSO, CCPSO and FIPS. The corresponding results are reported in Table II. These results are the best solutions taken from 10 independent runs of each algorithm. From the values of $J_{min}$, it can be observed that the optimal FOPID is the best controller structure followed by the optimal PID controller. The robust controllers (both FOPID and PID) have higher values of $J_{min}$ than the optimal controllers. This is in-line with our intuitive understanding, since the $J_{min}$ for the robust cases is the expected value over a range of controller gains and orders around the nominal value. For the robust designs, the CCPSO gives the best controller for the FOPID structure and the CPSO gives the best controller for the PID structure.

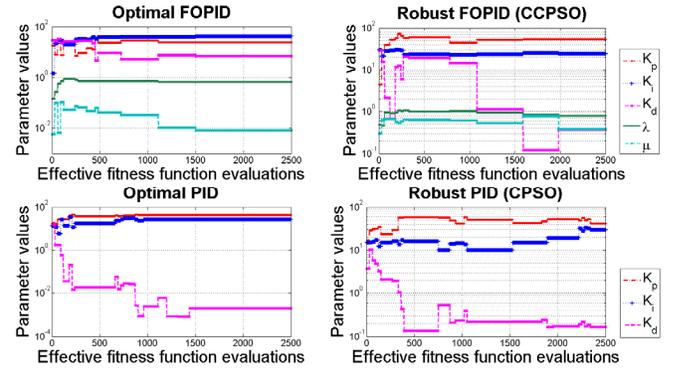

Fig. 7 Evolution of the PID/FOPID controller parameters or tuning knobs.

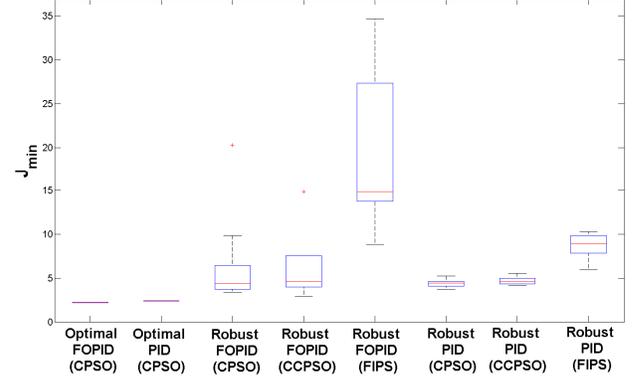

Fig. 8 Box and whisker plots for the $J_{min}$ for 10 runs of the PSO algorithms.

The convergence characteristics of the best found solutions for the four cases (optimal PID/FOPID and robust PID/FOPID) are reported in Fig. 6 and the corresponding evolution of the controller gains and orders are reported in Fig. 7. Both the values of $J_{min}$ and the controller parameters become almost constant towards the end of 2500 effective fitness function evaluations, indicating that the algorithms have converged to their respective minima.

Fig. 8 shows the statistical dispersion of the values of $J_{min}$ obtained from 10 independent runs of each algorithm. It can be seen that the algorithms converge to nearly the same value of $J_{min}$ for the optimal designs unlike those of the robust designs. Among the robust designs the variability in the FOPID controller is much higher than the PID controller. This can be attributed to the fact that higher degrees of freedom in the FOPID controller make many feasible solutions and therefore introduce multiple local minima in the objective function. The CPSO and the CCPSO algorithm work well for all the cases, while the FIPS does not give good solutions.

Also, traditional robust optimization techniques use some variants of convex optimization like linear matrix inequalities (LMIs), semi-definite programming etc. These have the advantage of guaranteed convergence to the global optima, but the objective function has to be convex in nature. However, in the present work the models contain significant nonlinearity (in the form of generation rate constraints) along with stochastic network induced delays which make the problem non-convex. These realistic effects make the optimization intractable using the traditional convex optimization techniques. So we resorted to a PSO based method which solves the optimization based control problem. But since the problem is inherently stochastic in nature, it does not give guaranteed convergence to the global optima. The PSO algorithm was therefore run multiple times and the box plots for the final optimal solutions are shown in Fig. 8.

TABLE II
PSO BASED TUNING OF PID AND FOPID CONTROLLERS

| Controller | Algorithm | Controller parameters | | | | | |
|---|---|---|---|---|---|---|---|
| | | $J_{min}$ | $K_p$ | $K_i$ | $K_d$ | $\lambda$ | $\mu$ |
| Optimal FOPID | CPSO | 2.17 | 24.181 | 41.567 | 6.863 | 0.675 | 0.008 |
| Optimal PID | CPSO | 2.39 | 42.599 | 27.536 | 0.002 | - | - |
| Robust FOPID | CPSO | 3.33 | 42.468 | 16.565 | 1.072 | 1.024 | 0.316 |
| | CCPSO | 2.96 | 53.795 | 24.054 | 0.364 | 0.799 | 0.378 |
| | FIPS | 8.84 | 25.413 | 14.936 | 7.948 | 0.953 | 0.211 |
| Robust PID | CPSO | 3.73 | 41.726 | 29.731 | 0.165 | - | - |
| | CCPSO | 4.17 | 43.973 | 29.363 | 0.213 | - | - |
| | FIPS | 5.95 | 24.436 | 16.207 | 0.255 | - | - |

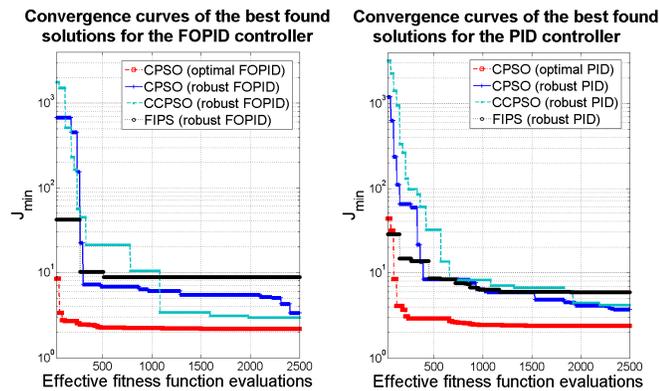

Fig. 6 Convergence characteristics of the PSO algorithm for PID/FOPID.

Fig. 9 shows the frequency deviation and control signal for the best optimal and robust designs of the FOPID and PID controller. Even though the optimal controllers are slightly better under nominal conditions, the robust controllers have good performance as well. Fig. 10 shows the individual powers, of the different components of the hybrid power system, for these solutions where it can be seen that there is not much appreciable difference between the optimal and the robust solutions. However unlike the robust solutions, the optimal solutions are not as good as at maintaining good performance in the presence of controller parameter uncertainties.

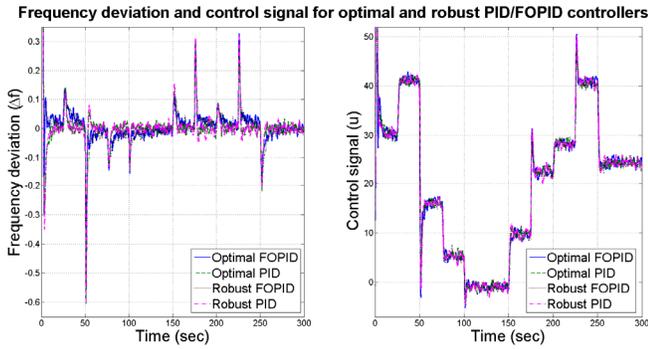

**Frequency deviation and control signal for optimal and robust PID/FOPID controllers**

Fig. 9 Performance of the optimal/robust PID and FOPID controllers.

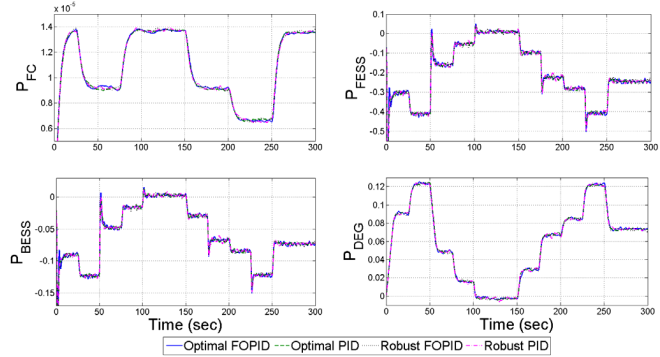

Fig. 10 Powers of the various components of the hybrid power system.

### B. Complexity of Robust Optimization Algorithms

The archive mechanism was implemented in the PSO algorithms to reduce the number of calls to the original function. Fig. 11 shows the evolution of the archive for the best obtained robust solution of the FOPID and PID controller. The number of elements in the archive does not exceed the maximum of 5000 as set during the initialization.

Fig. 12 shows a comparison of the effective fitness function evaluations (as given in (36)) for the robust PSO algorithms which reflects the large reduction in computational expense. During run-time, many of the solutions are evaluated by interpolating from the archive, so the actual fitness function evaluations (which require calls to the original computationally expensive objective function) are smaller for all the algorithms which implement the archive strategy. For the archive based robust algorithms, the number of actual function evaluations are less than 22% (as compared to the algorithms without the archive strategy) for the FOPID controller and less than 9% for the PID controller. This represents a huge saving in the computational budget for such robust optimization algorithms. It can also be observed that the FIPS has the least number of actual fitness function evaluations and is therefore much faster than the other two variants (CPSO and CCPSO). However, this also comes at the cost of lower performance as can be observed in Fig. 8, showing the box plots.

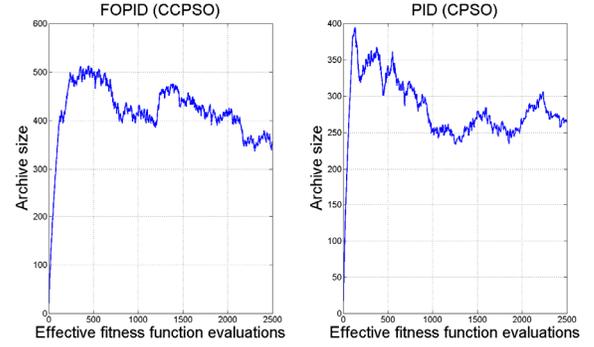

Fig. 11 Evolution of the archive size for the best obtained robust solution.

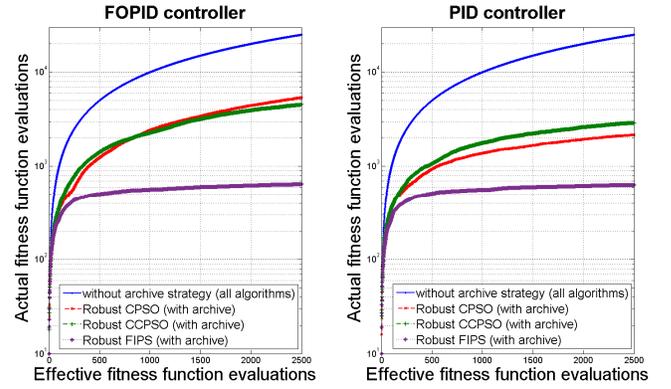

Fig. 12 Comparison of fitness function evaluations for the archive based robust PSO algorithms (corresponding to the best found solutions).

The simulations were run on a 64 bit Windows desktop with 16 GB memory and an Intel I7, 3.4 GHz processor. One function call takes ~0.5 min to run. Therefore the effective fitness function calculation according to eqn. (39) is $0.5 \times 10 = 5$ mins. All the algorithms were run for 2500 effective fitness function evaluations. Therefore the total time for each robust optimization without any archive strategy is $2500 \times 5 = 12500$ mins i.e. ~8.68 days. The archive strategy helps in obtaining the solutions using 22% (i.e. 1.9 days) of the expensive function evaluation (as compared to the previous case without the archive strategy) for FOPID and 9% of the expensive function evaluation (i.e. 0.78 days) for the PID controller.

### C. Performance Assessment under Perturbed Condition

To illustrate the effectiveness of robust optimization based controller design [23], 100 Monte-Carlo runs are conducted on both the optimal and the robust solutions by randomly perturbing the nominal values of the controller gains and orders. The ranges of perturbation are taken as $\Delta\{K_p, K_i, K_d, \lambda, \mu\} \in [10, 10, 0.15, 0.2, 0.2]$ (in both positive and the negative directions). The controller parameters are selected in this range using a uniform distribution and a Gaussian distribution and the statistical measures of $J$ (25) are reported in Table III. If the solutions become unstable or are infeasible (negative gain and order), a high value of $J = 10^4$ is assigned to those solutions. It can be observed, that with the perturbed controller parameters, the optimal controllers frequently become unstable/infeasible whereas, the robust designs are able to maintain acceptable performance near their nominal designs. Also the FOPID controller gives a better robust design than the PID controller as indicated by the expected value (mean) of the objective function $J$ in (25).

The present design focusses on maintaining acceptable performance when the controller parameters were perturbed. However, it is also important to assess whether the design can also handle uncertainty in the power system parameters. In Fig. 13 and Fig. 14, the grid frequency deviations have been shown for increase and decrease in the power system parameter $M$. Similar effects for perturbing other parameters ($D$, $K_{DEG}$, $T_{DEG}$, $K_{FESS}$, $T_{FESS} = T_{BESS}$, $K_{BESS}$, $\tau_{SC} = \tau_{CA}$) have been reported in the supplementary material. The change in controller performance ($J$) have also been reported in Table IV using the notation of % change = $(J - J_{nominal}) \times 100 / J_{nominal}$, where the perturbed cost function is compared and normalized against that with the nominal system parameters. Here, a positive % change indicates deterioration and a negative % change indicate improvement in control performance due to system parameter perturbation. From Fig. 13-14 it is evident that the FOPID scores over the PID, especially in terms of maintaining lower peaks in $\Delta f$. This can be verified from Table IV, for decrease in $M$ and increase in $K_{FESS}$, $\tau_{SC}$, $\tau_{CA}$.

TABLE III
STATISTICS OF THE OBJECTIVE FUNCTION ($J$) FOR 100 MONTE-CARLO RUNS

| Controller | $J$ for Uniform distribution | | $J$ for Gaussian distribution | |
|---|---|---|---|---|
| | Mean | Standard deviation | Mean | Standard deviation |
| Optimal FOPID | 5301.17 | 5014.88 | 4701.23 | 5014.97 |
| Optimal PID | 4301.63 | 4974.28 | 5001.33 | 5023.85 |
| Robust FOPID | *2.76* | *0.22* | *2.66* | *0.12* |
| Robust PID | 3.50 | 0.57 | 3.32 | 0.31 |

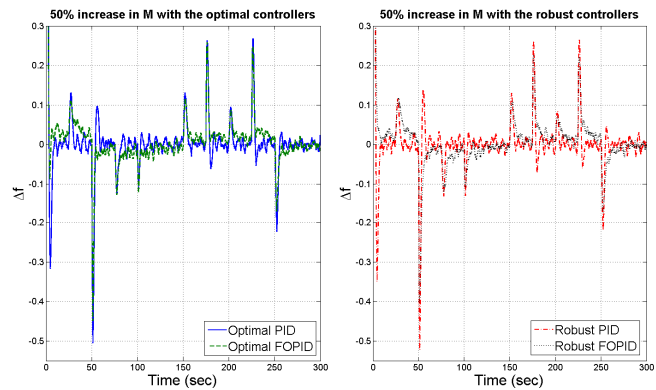

Fig. 13 Effect of increase in power system parameter $M$.

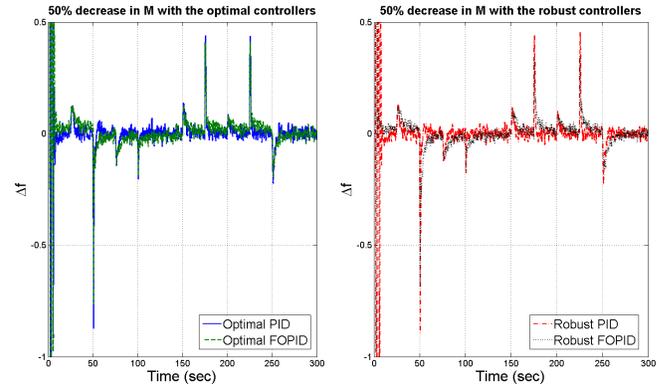

Fig. 14 Effect of decrease in power system parameter $M$.

TABLE IV
% CHANGE IN $J$ (25) FOR SYSTEM PARAMETER PERTURBATION

| System parameter | Perturbation | % change in $J$ | | | |
|---|---|---|---|---|---|
| | | Optimal FOPID | Optimal PID | Robust FOPID | Robust PID |
| $M$ | 50% increase | -31.628 | -25.718 | -37.341 | -31.913 |
| | 50% decrease | 132.354 | 417.819 | 35.368 | 614.078 |
| $D$ | 500% increase | -33.230 | -31.282 | -26.730 | -34.453 |
| | 500% decrease | 169.042 | 84.722 | 65.744 | 96.570 |
| $K_{DEG}$ | 500% increase | -0.963 | -2.226 | -1.933 | -2.673 |
| | 500% decrease | 32.341 | 5.565 | 26.459 | 1.507 |
| $T_{DEG}$ | 500% increase | 1.882 | -2.109 | 0.941 | -2.150 |
| | 500% decrease | 46.794 | 3.218 | 36.400 | -0.891 |
| $K_{FESS}$ | 70% increase | -54.423 | -10.649 | -63.984 | -0.417 |
| | 70% decrease | 187.858 | 211.266 | 141.587 | 172.016 |
| $T_{FESS} = T_{BESS}$ | 90% increase | 17.442 | 11.855 | 8.967 | 10.844 |
| | 90% decrease | -10.527 | -7.133 | -5.807 | -10.666 |
| $K_{BESS}$ | 70% increase | 2.136 | -2.685 | -1.977 | -3.095 |
| | 70% decrease | 17.652 | 21.726 | 11.305 | 13.353 |
| $\tau_{SC} = \tau_{CA}$ | 50% increase | 119.727 | 358.492 | 141.185 | 72.103 |
| | 50% decrease | -39.861 | -61.560 | -50.684 | -69.301 |

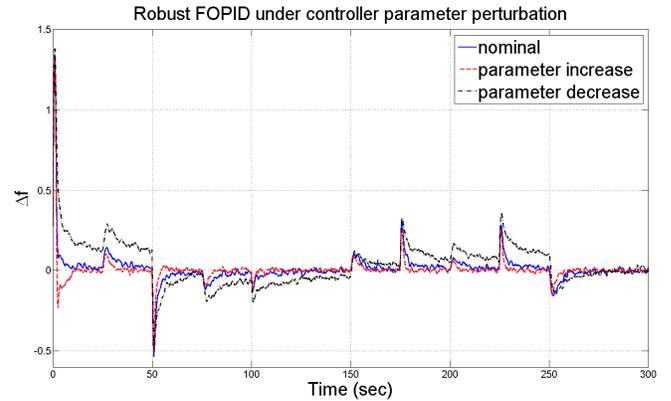

Fig. 15 Grid frequency deviation with robust FOPID under controller parameter perturbation.

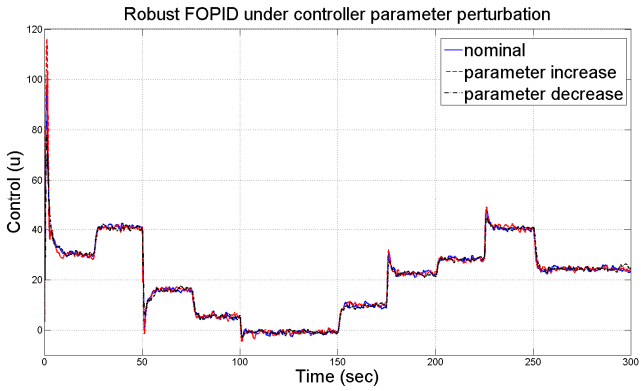

Fig. 16 Control signal with robust FOPID under controller parameter perturbation.

TABLE V
% CHANGE IN $J$ (25) FOR CONTROLLER PARAMETER PERTURBATION

| Controller | % increase | % decrease |
|---|---|---|
| Optimal FOPID | -48.606 | 20.712 |
| Optimal PID | 45.802 | 26.746 |
| Robust FOPID | 18.191 | 20.088 |
| Robust PID | 53.894 | -6.616 |

Next, the effect of variation in controller parameters, as summarized in Table III, is also shown graphically. Instead of random sampling from a distribution in the controller parameter space within the above mentioned range, the two extreme cases of maximum and minimum deviation of all the five FOPID tuning knobs (gains and orders) are considered here. The corresponding grid frequency deviation and control actions are depicted in Fig. 15 and Fig. 16 respectively for the robust FOPID. Similar results for the other three cases i.e. optimal/robust PID/FOPID are reported in the supplementary material. As evident, a decrease in the gains and integral order, make the system slower and takes more time to damp out the frequency oscillation. However, an improved performance is obtained by increasing the values of the PID/FOPID parameters. Although apparently it might seem that at the extreme ends i.e. at maximum and minimum values of the controller parameters the PID scores slightly better than the FOPID, but under random combination of the controller parameters, the PID controller often gives unstable/infeasible response as shown in the Monte Carlo runs in Table III. It is understandable from the fact that an increase in $K_p$, $K_d$ and/or $\mu$ increases oscillation and noise while a decrease $K_i$ and/or $\lambda$ produces poor steady-state behavior [6]. Therefore, increasing all the five FOPID parameters in either positive or negative direction may improve or deteriorate the performance which is explored in Table V. It is evident from Table V that with FOPID, the performance deterioration is less than that with the PID controller. Also, the robustness issue has been primarily considered in the controller parameter space during the design optimization. But the translation or propagation of the uncertainty from the controller parameter space to the system parameter space is not straight-forward as explored in [31] due to the nonlinearity and high complexity of the system. Our analysis shows that even under such design, the grid frequency oscillation is within acceptable limit considering system parameter perturbations.

### D. Discussions and Summary of Results

It is to be noted that in spite of having high system complexity like presence of the communication network in terms of stochastic delays before and after the controller along with large drift and smaller randomness in the renewable generations/load demand and rate-constraint nonlinearity in the energy storage elements, all the optimal controllers are successful in stabilizing the system. But the optimal solutions fail to perform in a similar manner under perturbed condition where the controller design with robust optimization score over the optimal design. It is also found that the FOPID outperforms the PID controller for both the optimal and the robust designs. The robust optimization based controller design is particularly important from the practical implementation of the FO elements in the FOPID, since they are realized using higher order analog/digital filters which have considerable variability in the frequency/time domain responses, depending on the approximation method and the order of realization. In spite of having several analytical stabilization results on simple linear networked control loops with stochastic delays [4], in the present context a simulation based design is adopted. This is because of the fact that due to the presence of stochastic forcing terms with sudden jumps (renewable generation, load fluctuation and random network induced delay) and the rate constraint nonlinearity in storage elements, such analytical fractional order controller design is difficult to achieve.

Also, it is found that there is hardly any variation in the performance of the optimal and robust controllers when the system is operating under nominal conditions. This is good because often there is a trade-off between the optimal and robust designs for FO controllers [25] and the robust controllers do not give good performance under nominal operating conditions. The focus of the paper is to show that the robust optimization based designs work well over their optimal counterparts when the controller parameters are perturbed (since during actual implementation, the FO controllers can be realized using various realizations and hence the actual gain and orders might be different from the designed ones). This is shown in Table III where it can be observed that the robust controllers give satisfactory performance when the controller parameters are perturbed. On contrary, the optimal controllers frequently become unstable or infeasible under such parameter perturbation. The FO controllers have been implemented in hardware for process control applications [14] and are also known to work better than their integer order counterparts in grid frequency oscillation damping problems [32, 33]. To the best of our knowledge, the FOPID controllers have not yet been validated in real time hardware demonstrator for smart grids or hybrid power systems. There have been several recent breakthrough in the hardware realization of FO controller e.g. digital realization using fixed and floating point representation [34], analog realization for fixed order [35] and finally variable selectable order [36]. This paper shows a proof-of-concept design methodology which facilitates the hardware implementation by making the system performance almost invariant to the type of realization used for the FO differ-integral operators in the FOPID controller. Future work can be aimed at physical hardware realization for the controller and real time implementation of the same on a smart grid or hybrid power system test bed.

In this paper, as an unreliable communication medium, we considered random time delays in the forward and feedback path. But there might be situations where a command is issued but never received by the device, which is known as packet dropout [7]. We have not modelled this phenomena in the present work, primarily because we expect the communication channel to be within a localized shared network and not over wide area networks or the internet where the packets travel through multiple routers and can be dropped due to buffer overflows. Nevertheless, the proposed robust optimization methodology is generic and would work even if we considered delays and packet dropouts together. Earlier investigations in [7] show that a FOPID controller, as also used here, is able to give better control performance over its integer order counterpart, in the presence of both stochastic network delays and packet dropouts. In the present study, a uniform distribution (i.e. a non-informative one) is considered for the stochastic delays, but actual measurements for the network delays can be done and a probability density function can be constructed for the same as shown in [8]. The proposed robust optimization methodology would work in both cases since it is generic in nature. It is also worthwhile to mention that recently there have been studies on controller design for distributed generation [37] and islanded microgrids [38] in a state space framework using $H_\infty$ theory and LMIs. These robust control theoretic approaches are only valid for a limited set of systems (mainly linear) with structured and unstructured uncertainties. The present robust simulation-optimization based control is intrinsically different from robust control theory and can handle nonlinear, stochastic systems with higher complexity. King *et al.* [39] recently proposed the use of machine learning techniques for algorithm selection in power system control, particularly in power flow management through active network management. A similar approach is adopted here as well, for algorithm selection in grid frequency control problem with distributed generation from a pool of four class of algorithms – optimal, robust, PID and FOPID controllers.

Regarding the choice of robust optimization, the present study only focusses on PSO variants [30] since it is widely established as a viable global optimizer. Many new algorithms like gravitational search, bat algorithm, glowworm swarm etc. have recently been developed and found to give better performance on test bench functions and some real world optimization problems. However, the performance of the robust version of these optimization algorithms have not been investigated and different strategies to reduce the simulation runs e.g. the archive strategy [28] as used in the present work, have not yet been developed for these algorithms. The methodology presented in the paper would be valid with any robust optimization algorithm and different popular variants of PSO have been used in the present study to demonstrate plausible simulation results.

## VI. CONCLUSIONS

The robust optimization based tuning algorithms (involving CPSO variants) are employed for FOPID controllers to handle a distributed energy generation system with rate constraint nonlinearity, over an unreliable communication network. The optimal designs give better performance than their robust counterparts for the nominal values of the controller parameters, but the performance severely degrades if the controller parameters are perturbed from their optimal values. The robust designs for both the PID/FOPID controllers are tested for uncertainties in the controller parameters as well as the power system parameters. Future work may be directed towards looking at analytical controller designs methods for such complex nonlinear and interconnected power systems.

## APPENDIX

Additional analysis of robustness and high resolution enlarged images are available in the supplementary material.

# Supplementary Material

### Indranil Pan and Saptarshi Das

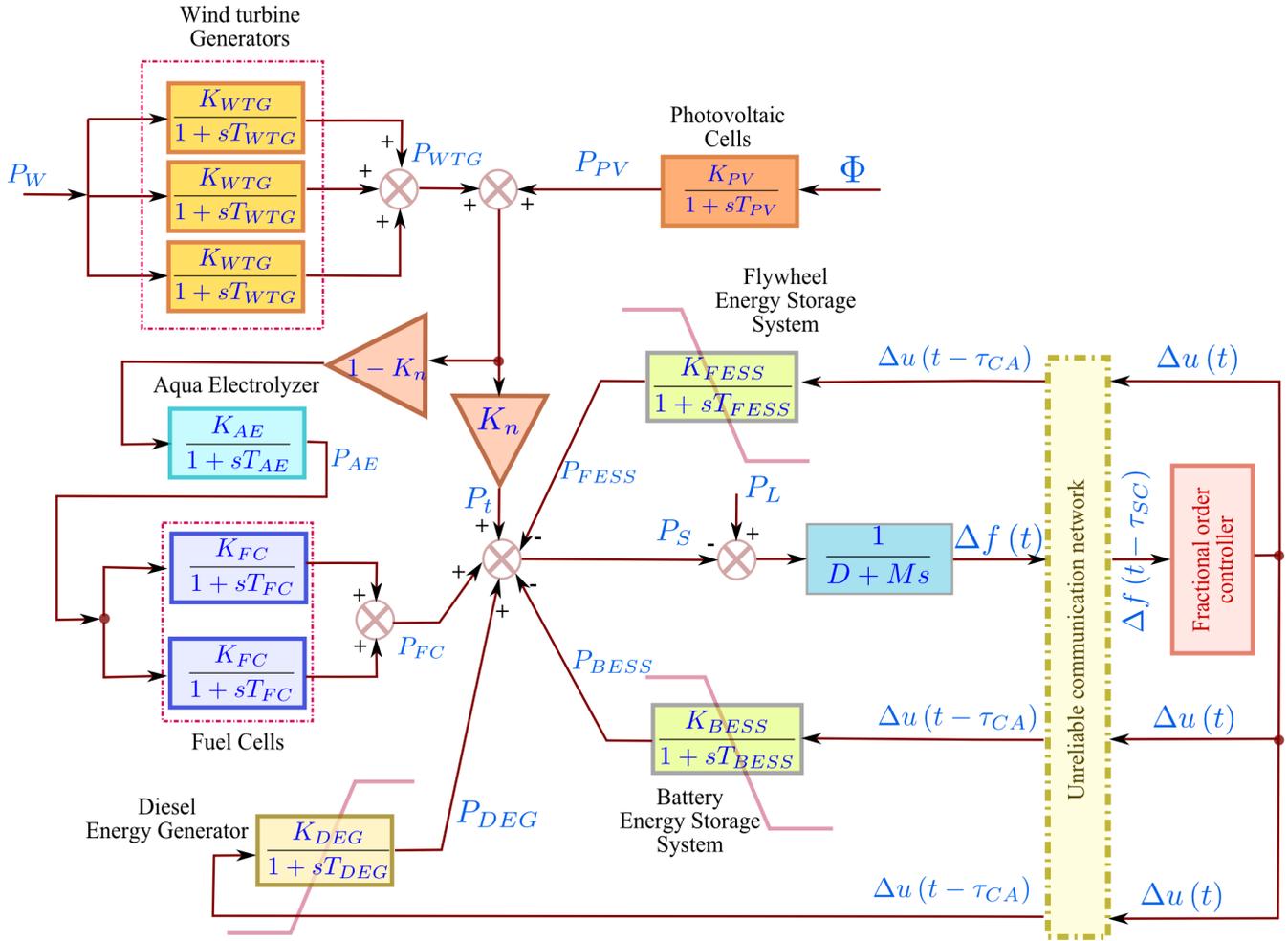

Fig. 1 Schematic of the hybrid power system used in this study.

I. Pan is with the Department of Earth Science and Engineering, Imperial College London, Exhibition Road, SW7 2AZ, UK (email: i.pan11@imperial.ac.uk).
S. Das is with the School of Electronics and Computer Science, University of Southampton, Southampton SO17 1BJ, UK (e-mail: s.das@soton.ac.uk).

**Stochastic fluctuation of the power generated by renewable sources and demand load**

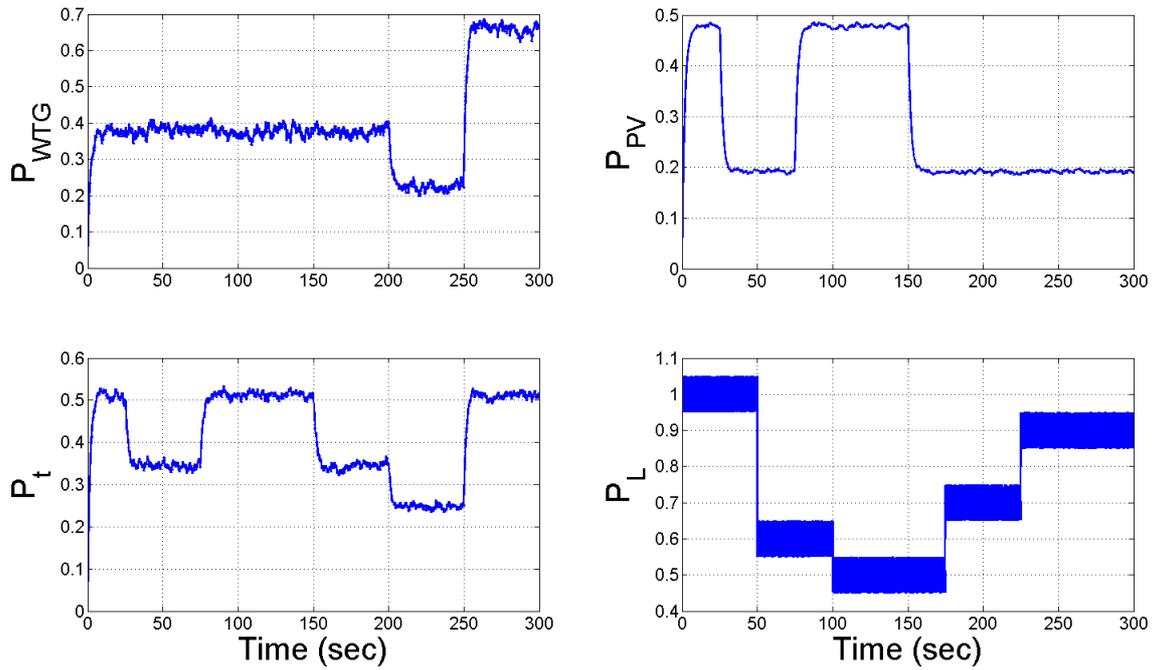

Fig. 2 A single realization of the renewable generations and demand powers which are independent of the controller structure.

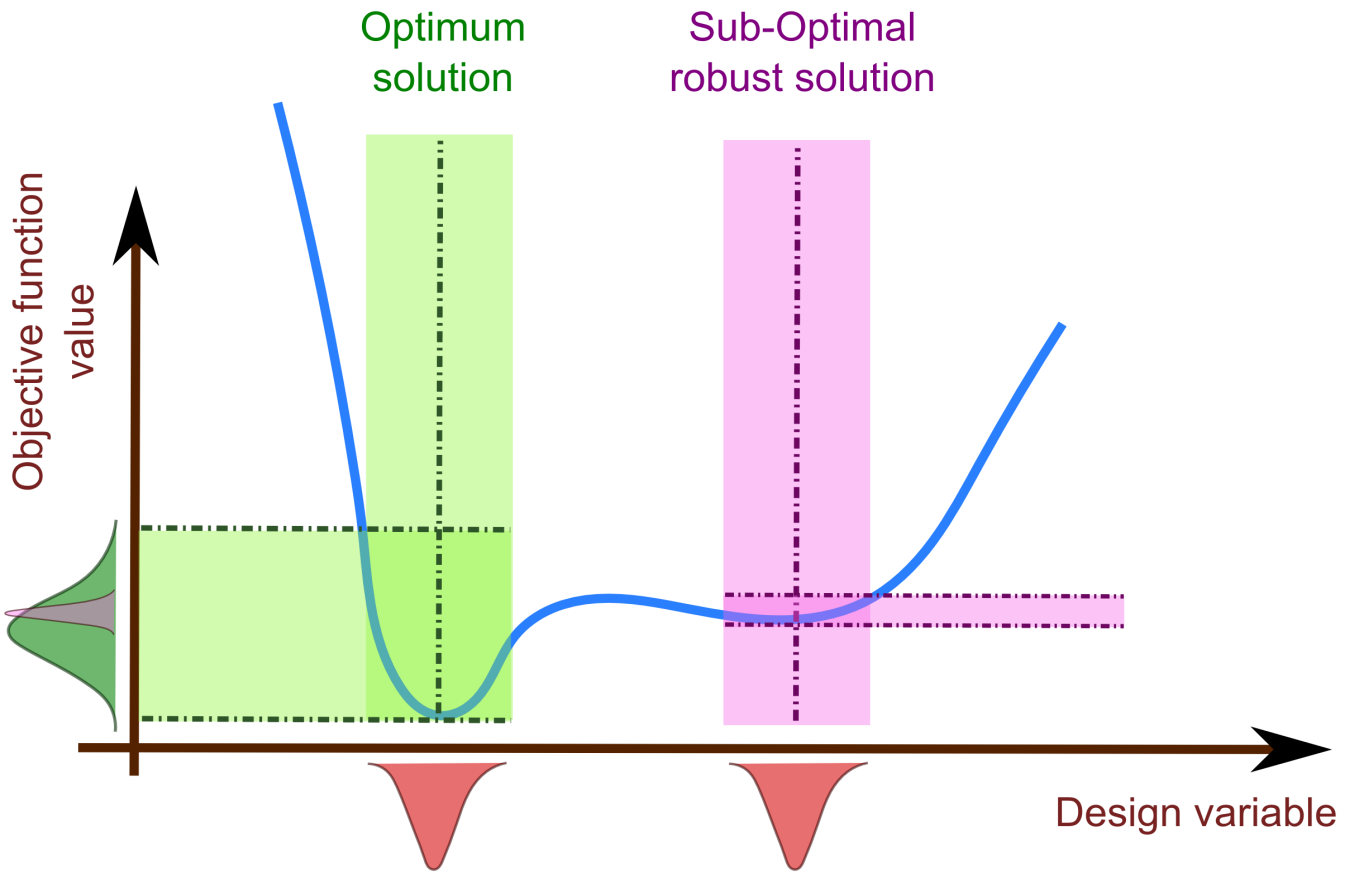

Fig. 3 Schematic showing the difference between an optimum solution and a robust solution.

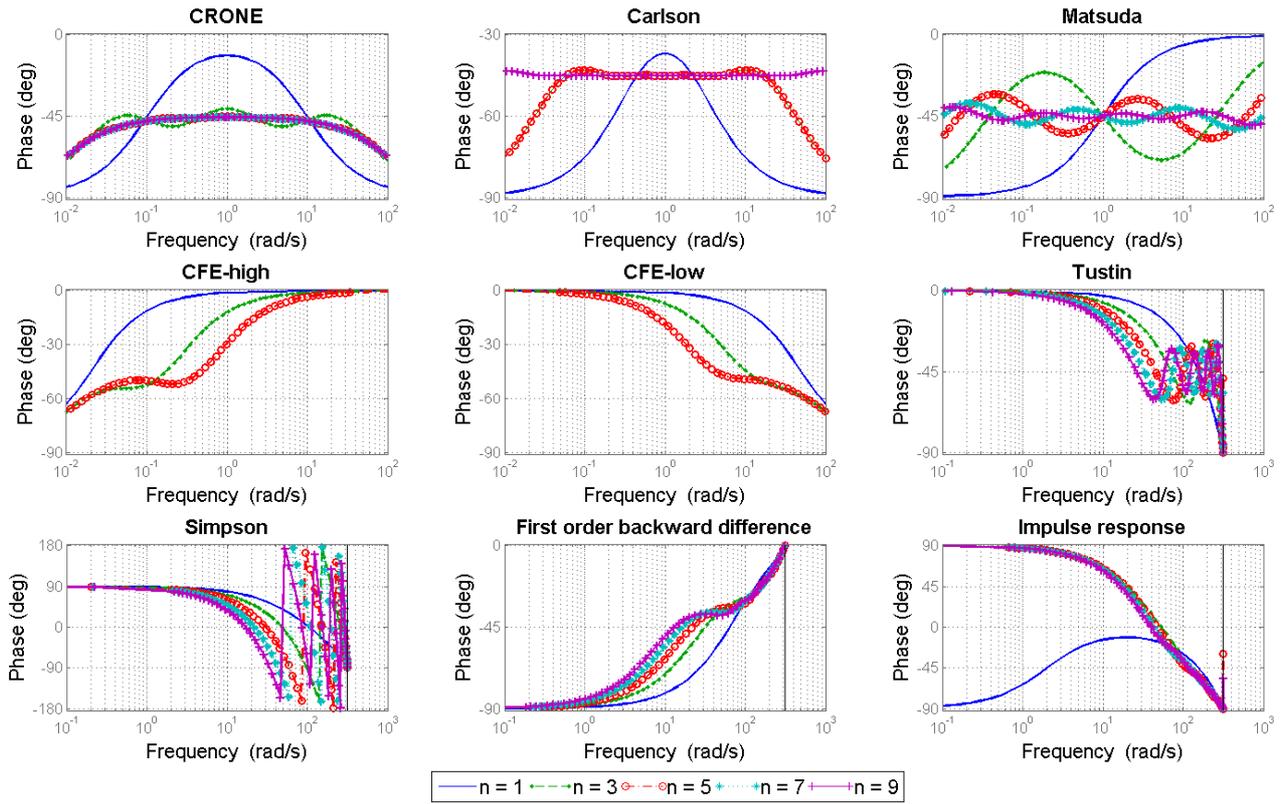

Fig. 4 Phase responses of different band-limited realizations of FO element $1/\sqrt{s}$ with different methods and order of approximation.

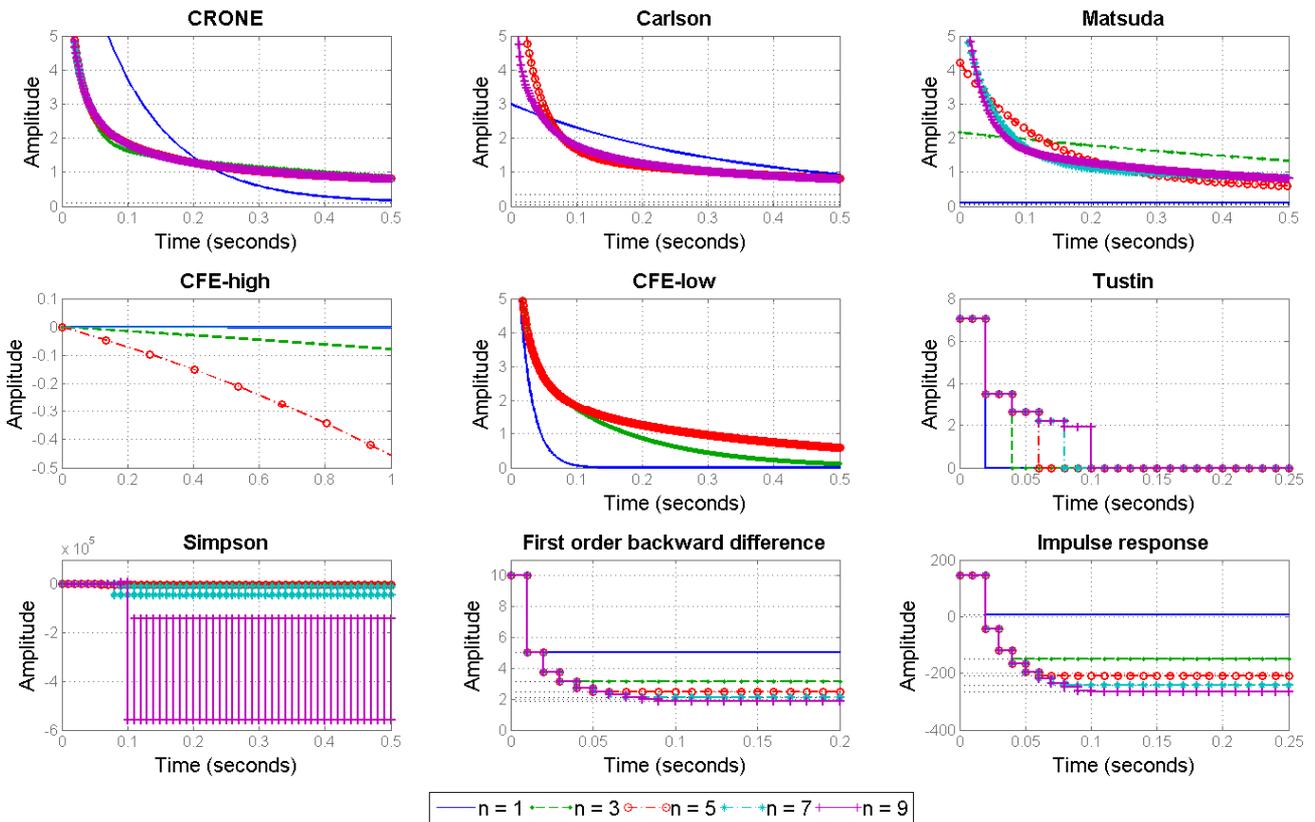

Fig. 5 Impulse responses of FO element $1/\sqrt{s}$ with different methods/orders.

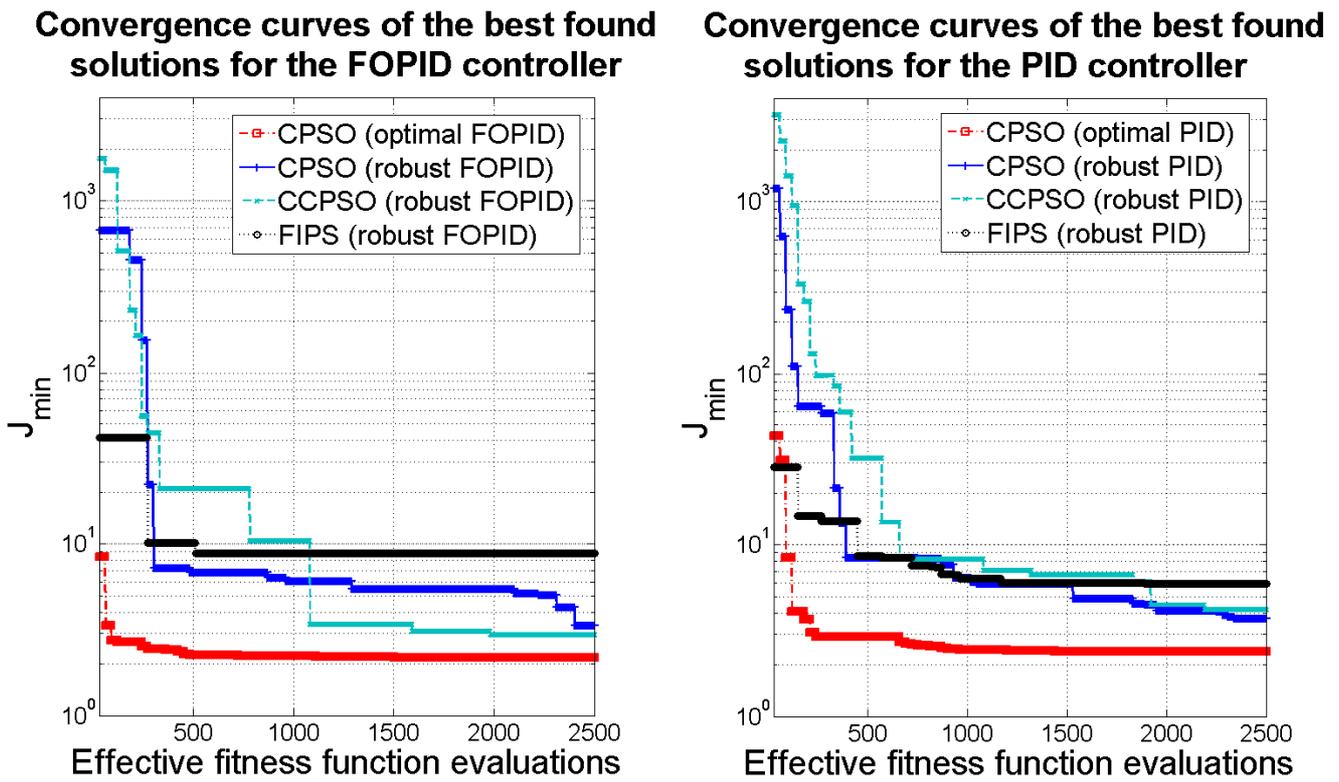

Fig. 6 Convergence characteristics of the PSO algorithm for PID/FOPID.

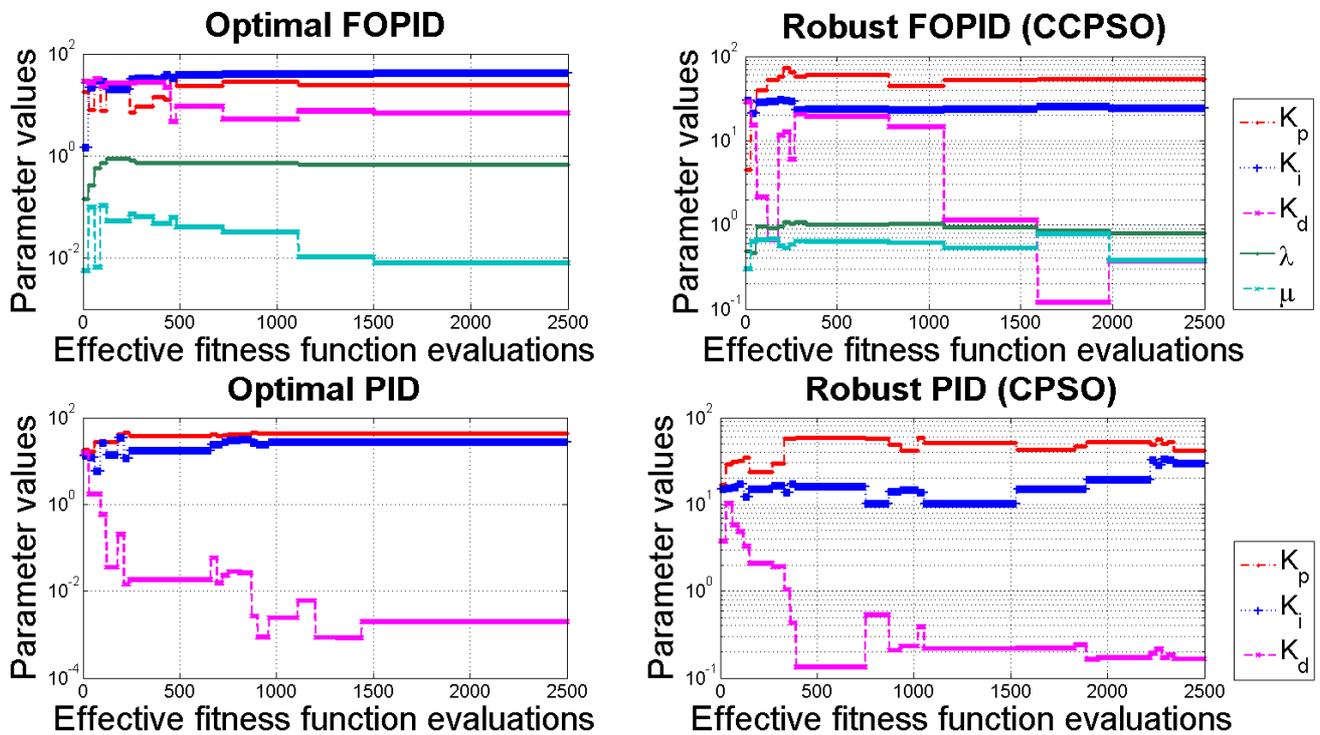

Fig. 7 Evolution of the PID/FOPID controller parameters or tuning knobs.

**Box and whisker plot for the final optimized values in 10 runs of each algorithm**

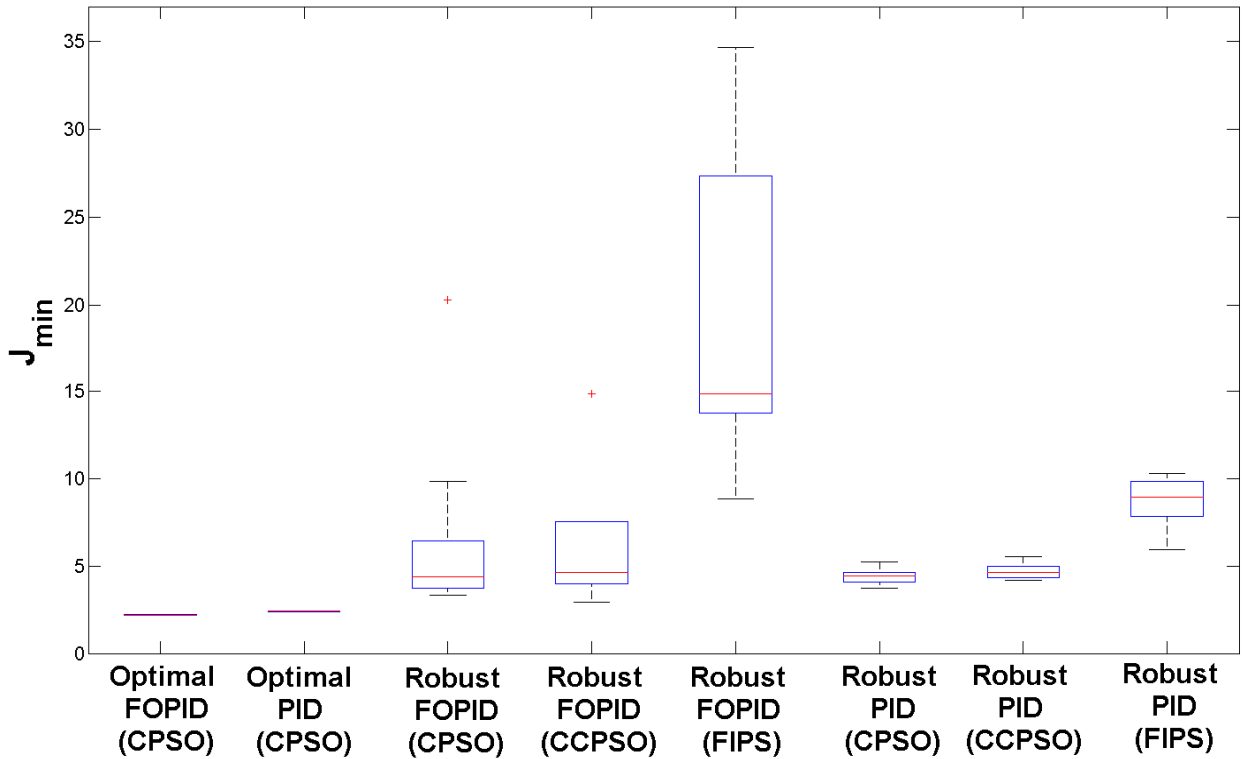

Fig. 8 Box and whisker plots for the $J_{min}$ for 10 runs of the PSO algorithms.

**Frequency deviation and control signal for optimal and robust PID/FOPID controllers**

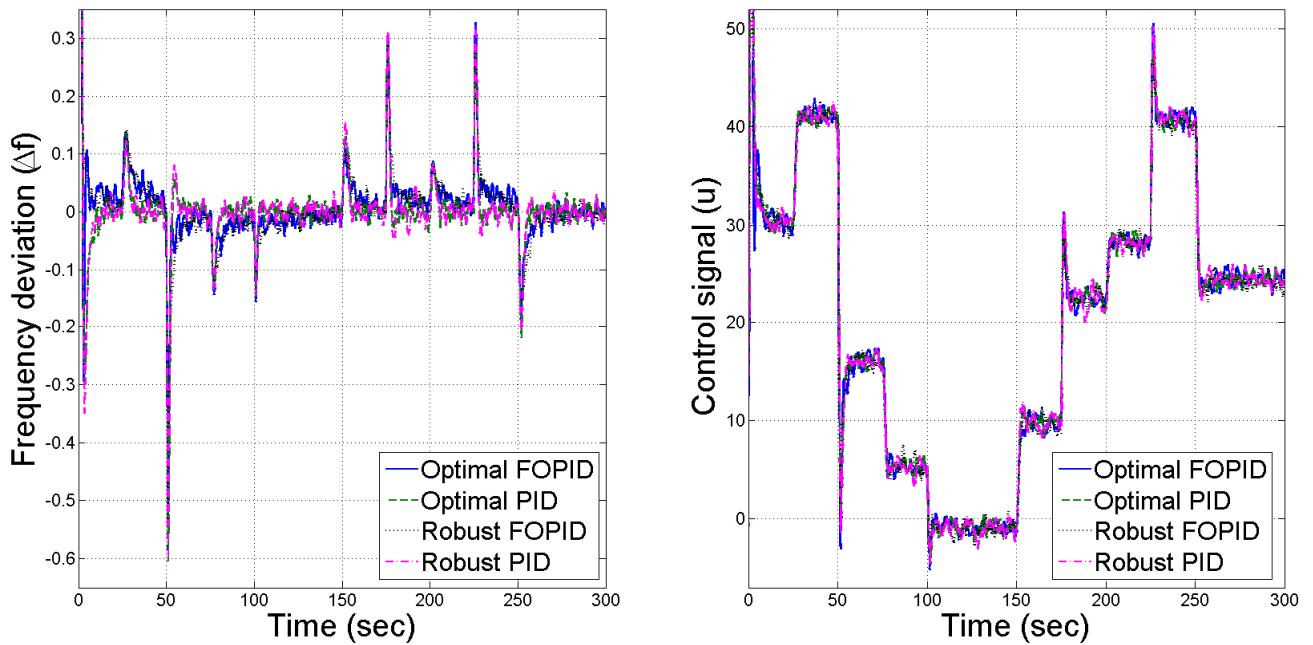

Fig. 9 Performance of the optimal/robust PID and FOPID controllers.

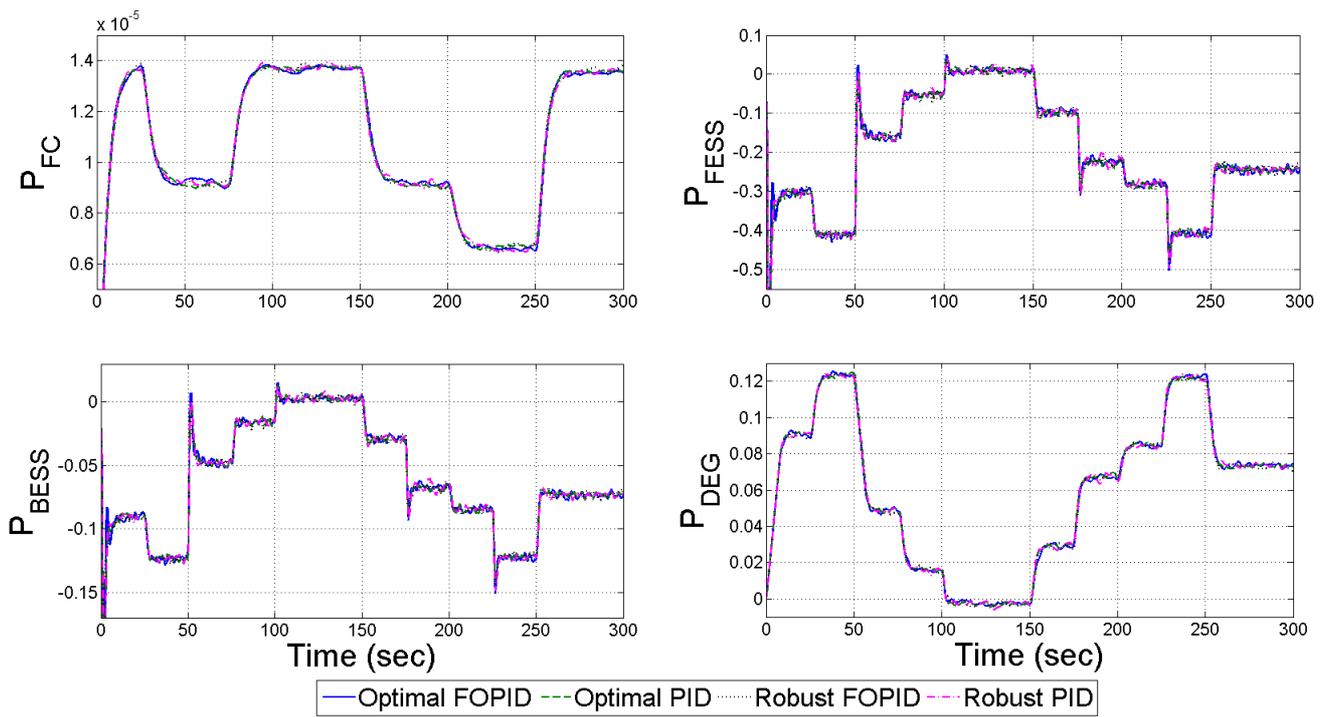

Fig. 10 Powers of the various components of the hybrid power system.

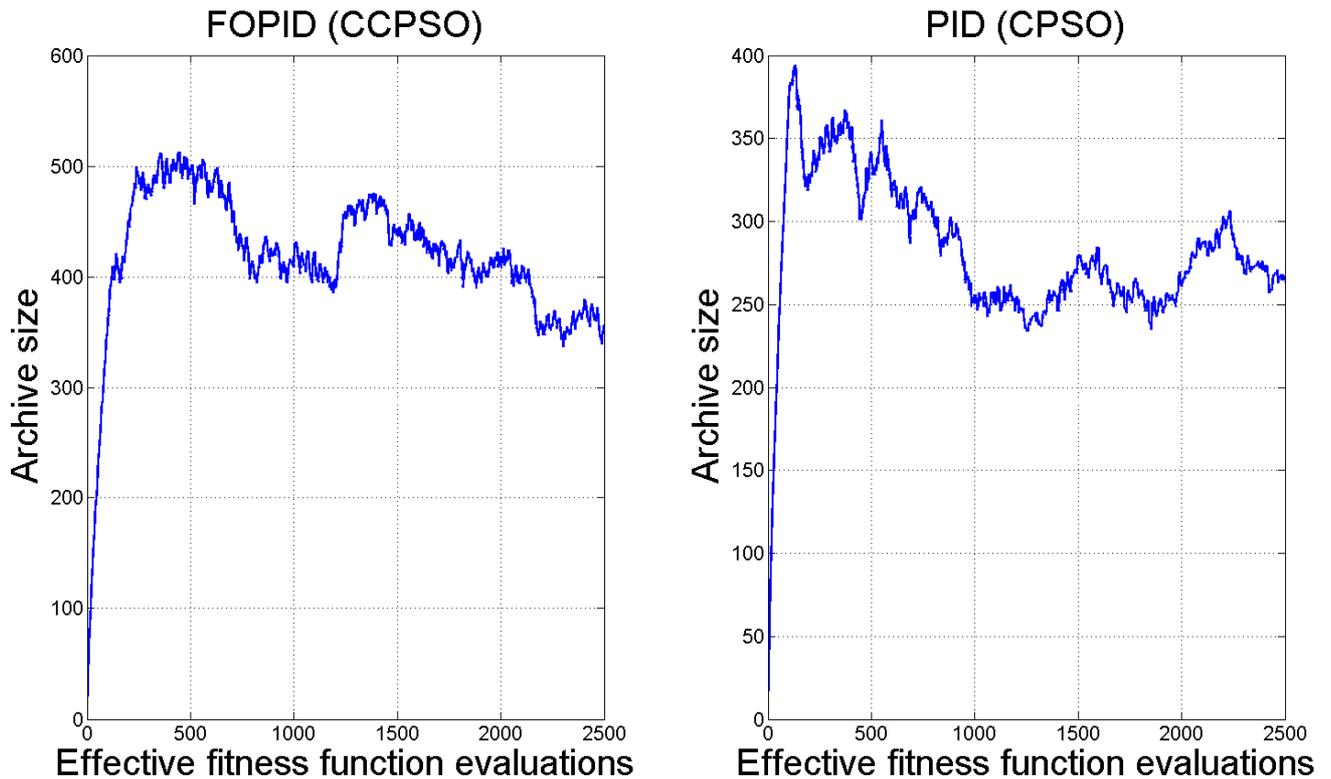

Fig. 11 Evolution of the archive size for the best obtained robust solution.

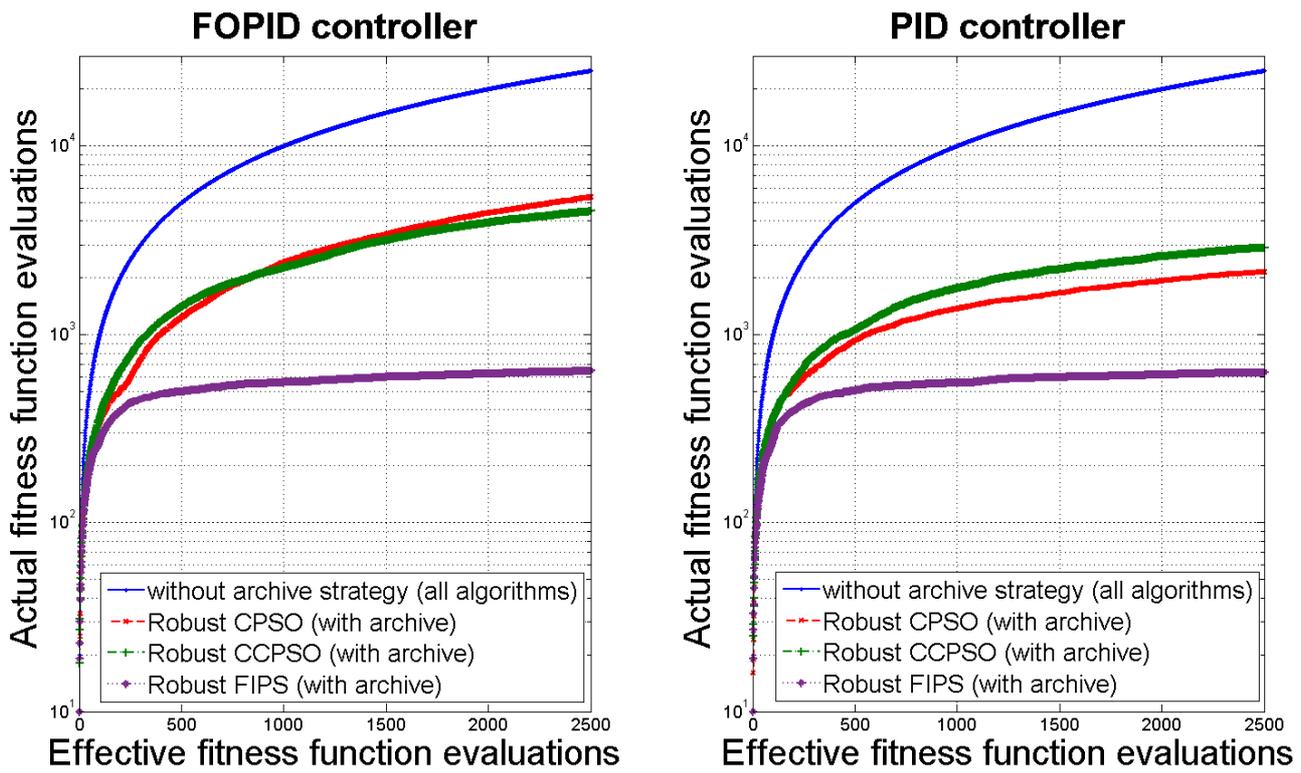

Fig. 12 Comparison of fitness function evaluations for the archive based robust PSO algorithms (corresponding to the best found solutions).

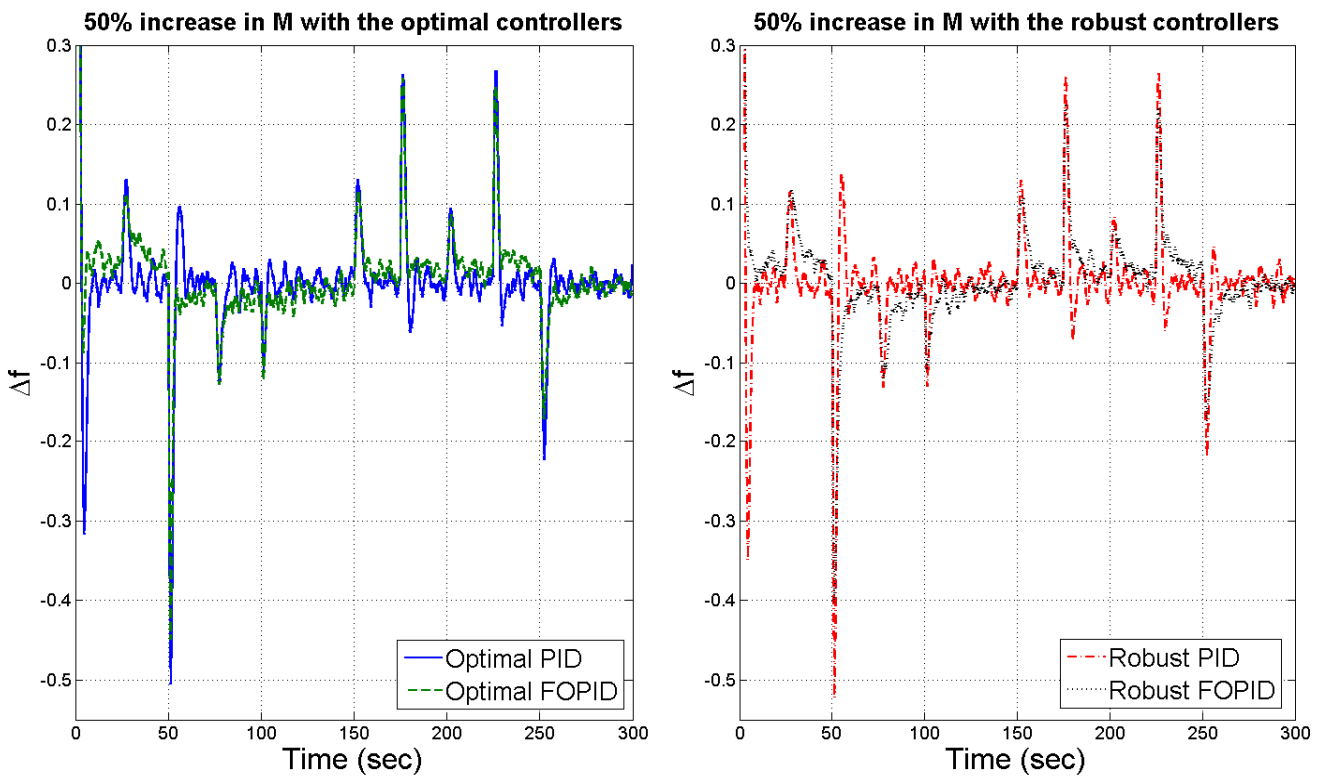

Fig. 13 Effect of increase in power system parameter *M*.

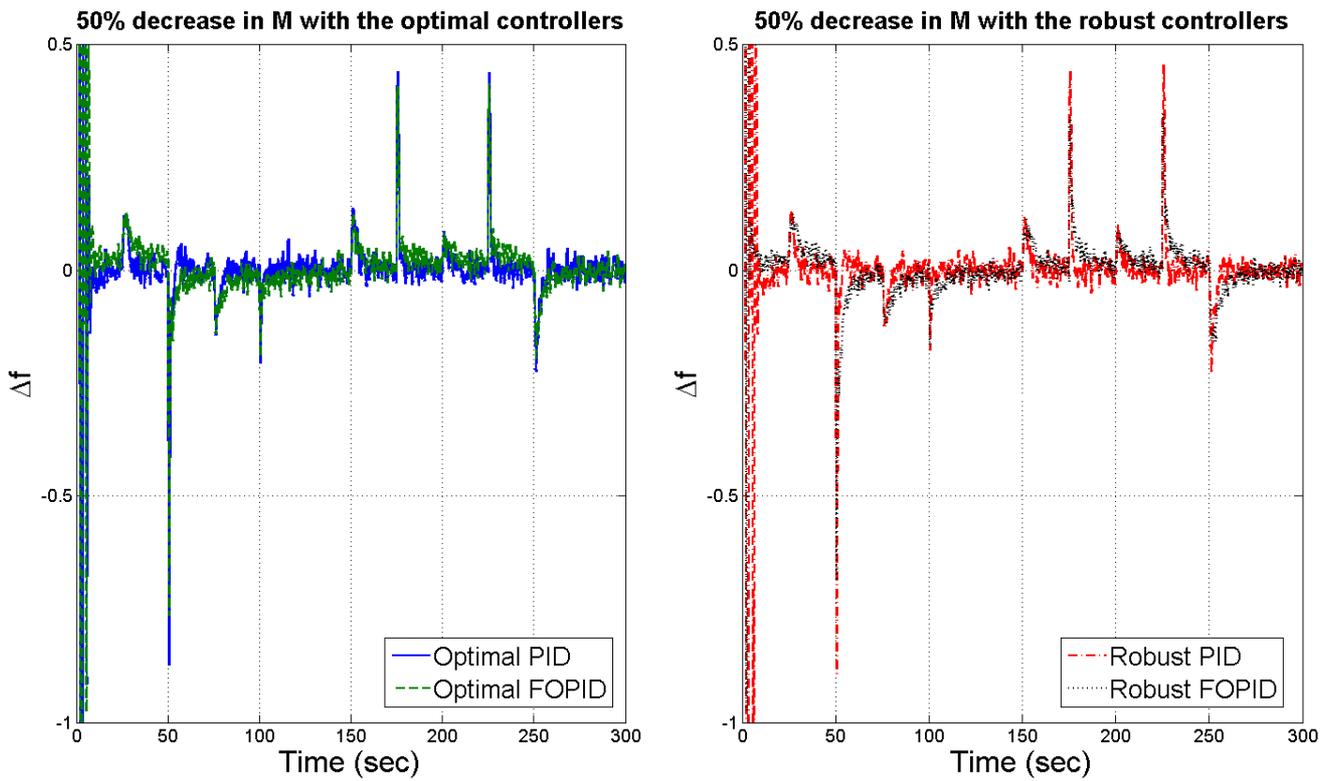

Fig. 14 Effect of decrease in power system parameter *M*.

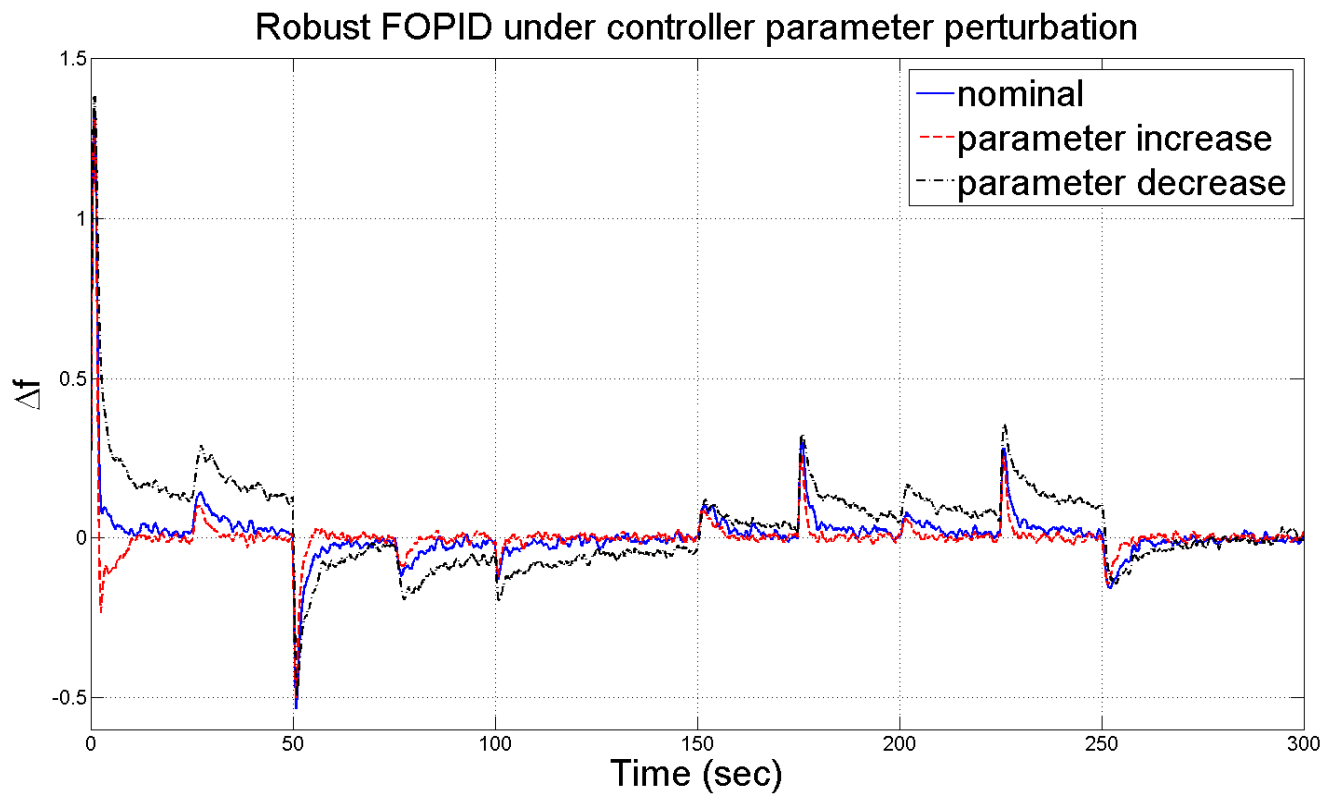

Fig. 15 Grid frequency deviation with robust FOPID under controller parameter perturbation.

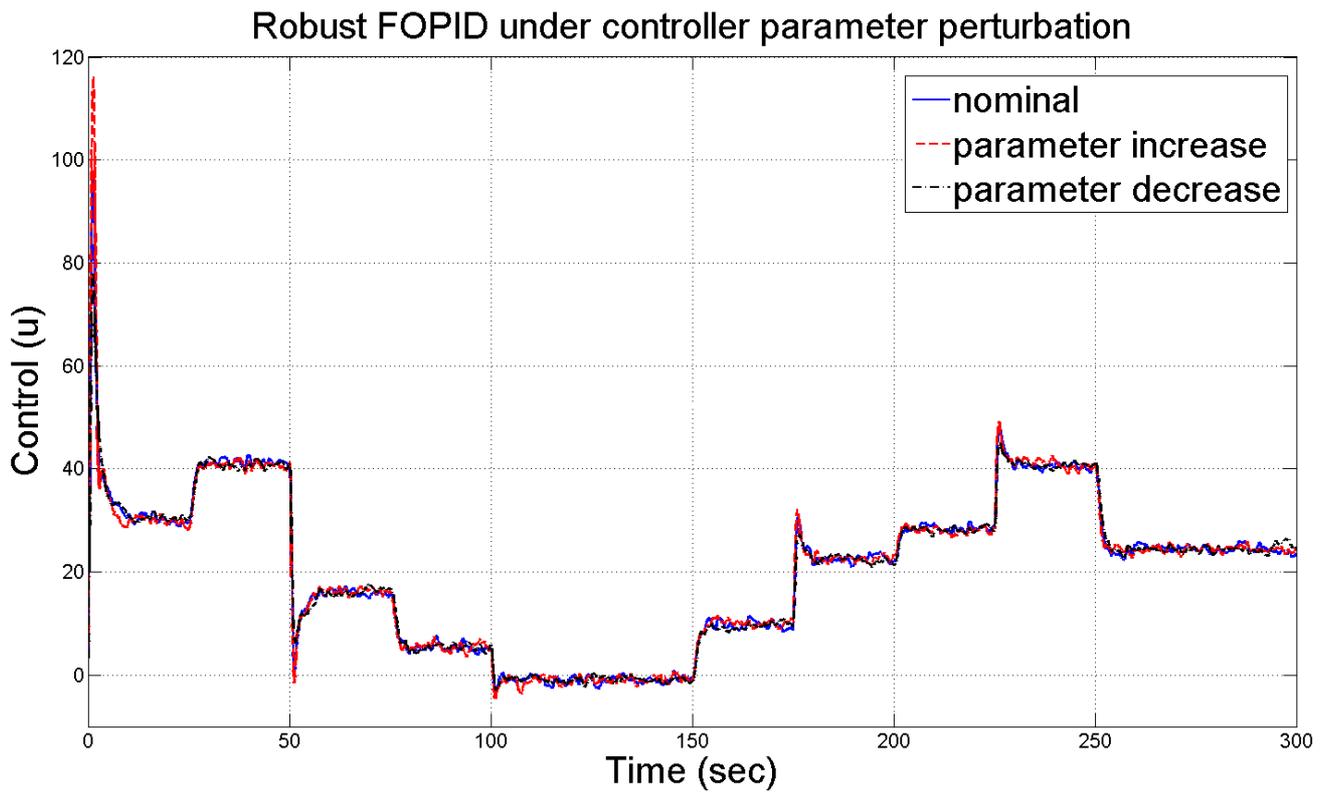

Fig. 16 Control signal with robust FOPID under controller parameter perturbation.

Grid frequency oscillations for perturbation in other controller parameters similar to $M$ (Fig. 13 and Fig. 14) like $D$, $K_{DEG}$, $T_{DEG}$, $K_{FESS}$, $T_{FESS} = T_{BESS}$, $K_{BESS}$, $\tau_{SC} = \tau_{CA}$ are shown in Fig. 17 and Fig. 18 respectively.

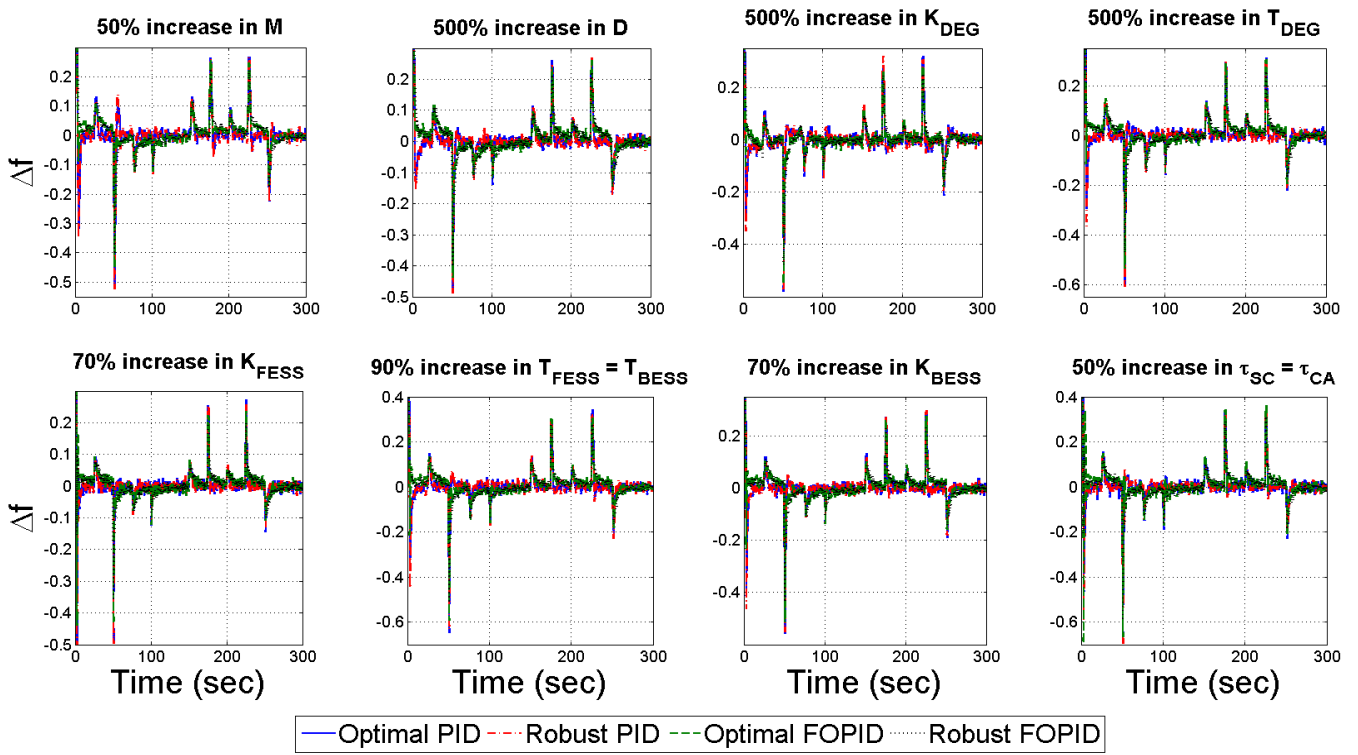

Fig. 17 Effect of increase in power system parameters.

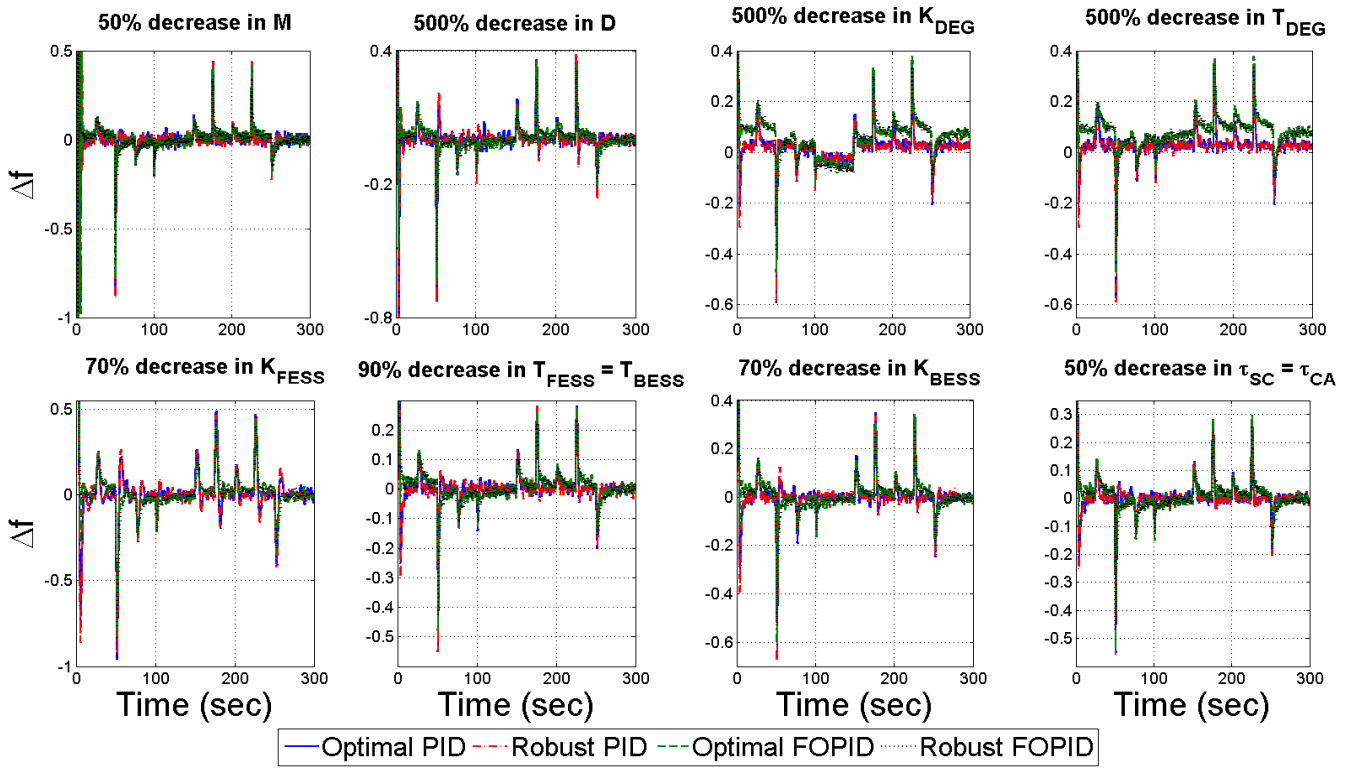

Fig. 18 Effect of decrease in power system parameters.

Similar to the case of robust FOPID (Fig. 15 and Fig. 16), the effect of parameter perturbation for all the four controller structures (optimal/robust, PID/FOPID) are shown in Fig. 19 and Fig. 20 respectively.

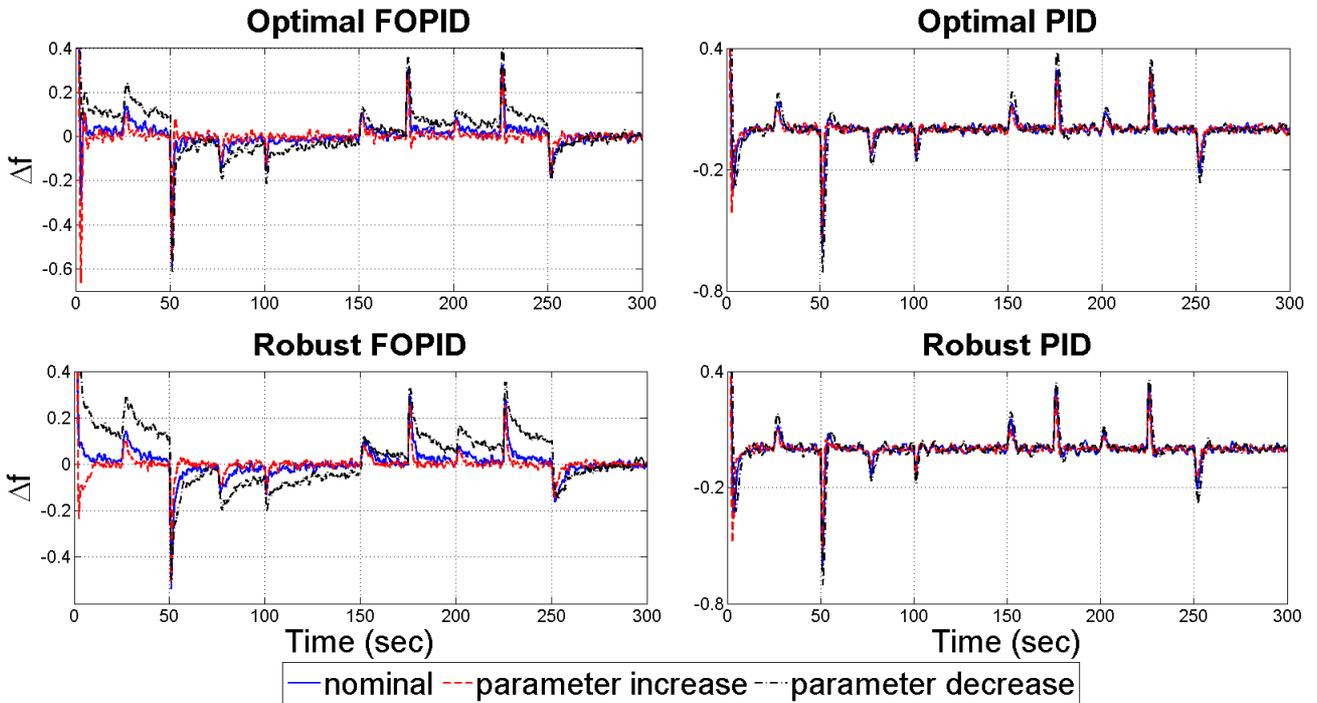

Fig. 19 Grid frequency deviation under controller parameter perturbation.

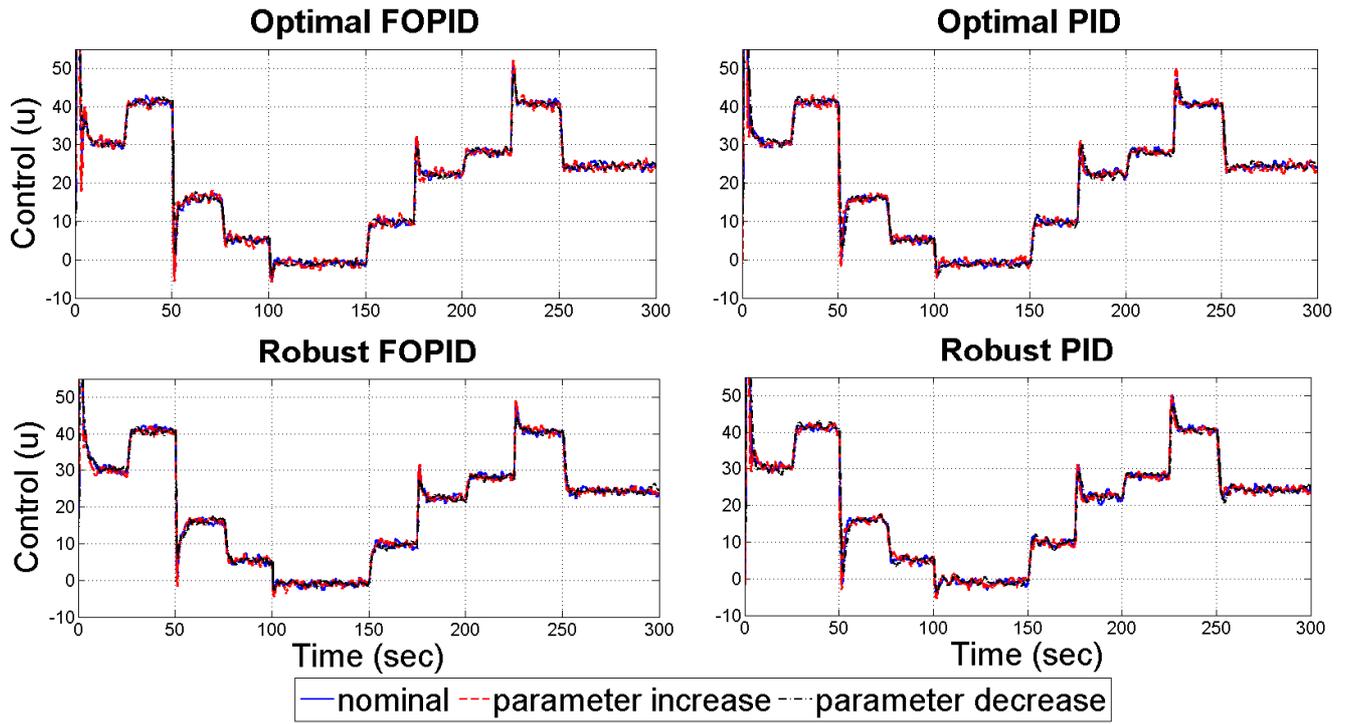

Fig. 20 Control signal under controller parameter perturbation.